\documentclass [article]{aa}
\usepackage{graphicx}
\usepackage{txfonts}

 \usepackage[T1]{fontenc}
 \usepackage{textcomp}
 \usepackage{mathptmx}
 \usepackage{courier}
 \usepackage[scaled=0.94]{helvet}
 \usepackage{hyperref}
 \hypersetup{
     final=true,
     pageanchor=true,
     hyperfigures=false,
     colorlinks=true,
     breaklinks=true,
     linkcolor=blue,
     citecolor=blue,
     urlcolor=blue,
     pdfpagemode=UseNone,
     pdfpagelayout=OneColumn,
     pdfview=FitBH,
     pdftitle={Analysis of the variation of physical quantities of granular structures},
     pdfauthor={Puschmann et al.},
     pdfsubject={Solar Physics}}

\begin{document}

\title{Time series of high resolution photospheric spectra in a quiet region of the sun}
\subtitle{II. Analysis of the variation of physical quantities of granular structures}

\author{K. G. Puschmann \inst{1,2,3} \and B. Ruiz Cobo \inst{2} \and M. V\'azquez \inst{2} \and J. A. Bonet \inst{2} \and A. Hanslmeier\inst{3}}

\offprints{K. G. Puschmann}

\institute{Universit\"ats-Sternwarte, Geismarlandstr. 11, D-37083 G\"ottingen, Germany\\
\email{kgp@uni-sw.gwdg.de}
\and
Instituto de Astrof\'\i sica de Canarias, C/V\'\i a 
L\'actea s/n,E-38200 La Laguna, Spain\\
\email{brc@ll.iac.es,  jab@ll.iac.es, mva@ll.iac.es}
\and 
Institut f\"ur Geophysik, Astronomie und Metereologie, Universit\"at Graz, Universit\"atsplatz 5, A-8010 Graz, Austria\\
\email{arnold.hanslmeier@kfunigraz.ac.at}}

\date{Received date; accepted date}
%\maketitle
%\markboth {}{}
           
\abstract{
From the inversion of a time series of high resolution slit spectrograms obtained from the quiet sun, the spatial and temporal distribution of the thermodynamical quantities and the vertical flow velocity is derived as a function of optical depth ($\log\tau$) and geometrical height ($z$). Spatial coherence and phase shift analyses between temperature and vertical velocity depict the height variation of these physical quantities for structures of different size. An average granular cell model is presented, showing the granule--intergranular lane stratification of temperature, vertical velocity, gas pressure and density as a function of $\log\tau$ and $z$. Studies of a specific small and a specific large granular cell complement these results. A strong decay of the temperature fluctuations with increasing height together with a less efficient penetration of smaller cells is revealed. The $T$\,-\,$T$ coherence at all granular scales is broken already at $\log\tau$\,=\,$-$1 or $z$\,$\sim$\,170\,km. At the layers beyond, an inversion of the temperature contrast at granular scales $>$\,1\farcs5 is revealed, both in $\log\tau$ and $z$. At deeper layers the temperature sensitivity of the $H^-$ opacity leeds to much smaller temperature fluctuations at equal $\log\tau$ than at equal $z$, in concordance with Stein \& Nordlund (\cite{steinnordlund98}). Vertical velocities are in phase throughout the photosphere and penetrate into the highest layers under study. Velocities at the largest granular scales ($\sim$\,4\arcsec) are still found even at  $\log\tau$\,$\sim$\,$-$2.8 or $z$\,$\sim$\,370\,km. Again a less efficient height penetration  of smaller cells concerning convective velocities is revealed, although still at $\log\tau$\,$\sim$\,$-$2 or $z$\,$\sim$\,280\,km structures $>$\,1\farcs4 are detected. A similar size distribution of velocity and temperature structures with height provides observational evidence for substantial overshoot into the photosphere. At deep photospheric layers, the behaviour of the vertical velocities reflected in simulations is for the first time qualitatively reproduced by observations: Intergranular velocities are larger than the granular ones and, both reach extrema, where the granular one is shifted towards higher layers.

}

\titlerunning{Time series of high resolution photospheric spectra}
\maketitle
%
%_________________________________

\section{Introduction}
\label{inversion}

During the last two decades the vertical structure of the solar photosphere has been investigated by different methods. 

Statistical analyses of a few parameters like continuum intensity, line core intensity, velocities (derived from Doppler shift measurements in the line core), equivalent width and full width at half maximum are often used, in addition to line bisectors, in order to characterise the shape of photospheric lines and to infer important physical parameters.

Most investigations have been based on correlation and spectral analysis techniques applied to one-dimensional (1D) slit spectrograms (Nesis et al. \cite{nesis88}, \cite{nesis92}, \cite{nesis93}; Komm et al. \cite{komm90}, \cite{komm91a}, \cite{komm91b}, \cite{komm91c}; Hanslmeier et al. \cite{hanslmeier91a}, \cite{hanslmeier91b}, \cite{hanslmeier93}, \cite{hanslmeier94}, and references therein), but without filtering properly for oscillations, which strongly mask the convective signature.

Altrock et al. (\cite{altrock}) and Nesis et al. (\cite{nesis99}, \cite{nesis01}) followed the evolution of selected granules during short periods of time.
Keil (\cite{keil}) and Johanneson et al. (\cite{johanneson}) obtained 2D spectral images derived from 1D spectrograms by scanning a certain area of the solar surface. Collados et al. (\cite{collados}) obtained improved data using the Correlation Tracker developed at the Instituto de Astrof\'\i sica de Canarias.

2D spectrograms have been obtained using a Multichannel Subtractive Double Pass (MSDP) spectrograph (Roudier et al. \cite{roudier}; Espagnet et al. \cite{espagnet93}, \cite{espagnet95}) or Fabry-Perot interferometers (Salucci et al. \cite{salucci}; Bendlin \& Volkmer \cite{bendlin};  Krieg et al. \cite{krieg}; Hirzberger et al. \cite{hirzberger01}). 

On the other hand, one of the main issues of solar atmospheric studies is to determine in a more accurate way the stratification of physical quantities throughout the solar photosphere. This can be done by different techniques. 

One is to assume a certain model atmosphere and to obtain through forward modelling synthesised observable parameters which can directly be compared with observations. See to the numerical simulations of solar convection by Steffen et al. (\cite{steffen89}, \cite{steffen03}), Stein \& Nordlund (\cite{stein89}), Gadun et al. (\cite{gadun97}, \cite{gadun99}, \cite{gadun00}), Freytag et al. (\cite{freytag02}), and Wedemeyer et al. (\cite{wedemeyer04}). 

Another technique is to invert the observations in order to obtain a model atmosphere through iteratively comparing synthetic observables with real ones. Inversion methods address the problem of the determination of the stratification of physical quantities by simultaneously varying all (or at least as many as possible) relevant physical quantities in a parameterised atmospheric model in order to obtain iteratively the best fit to the observed line profiles.

The first semi-empiric 2D model of the solar atmosphere, including a physical model of granulation, was developed by Nelson (\cite{nelson}). Later Skumanich \& Lites (\cite{skumanich}) obtained a model using their inversion code based on the Milne-Eddington approximation. Several inversion methods have been developed at the IAC. Ruiz Cobo \& del Toro Iniesta (\cite{ruizcobo92}) presented the inversion code SIR (Stokes inversion based on response function), dealing with arbitrary stratifications of physical quantities in the photosphere. Although the SIR code was applied to deduce the structure of granules and intergranular lanes in the sun (Rodr\'{\i}guez Hidalgo et al. \cite{rodriguez96}, Ruiz Cobo et al. \cite{ruizcobo96}), the results have never been published in cited journals due to three fundamental criticisms: The data have not been filtered for oscillations, the spectral lines used for these studies have not been corrected for departures from LTE, and one of these lines shows strong magnetic sensitivity. A very similar code has been applied on low spatial resolution but high signal to noise ratio spectra (Frutiger et al. \cite{frutiger}). 

The state-of-the-art knowledge of the granular phenomenon can be described in the following way: Convective overshoot of the plasma from the solar convection zone into the photosphere forms a pattern of bright cellular elements showing upwards motion -- the granules -- surrounded by a network of dark intergranular lanes, where down-flow motions are observed.  According to correlation analyses (Deubner \cite{deubner88}; Salucci et al. \cite{salucci}; Espagnet et al. \cite{espagnet95}) the horizontal temperature fluctuations associated with these motions decrease rapidly with increasing height until they vanish. Only the largest granules ($>$\,1\farcs5) contribute to the brightness pattern observed above where the brightness contrast is inverted. Note that the values of the height where the temperature fluctuations vanish varies a lot in the literature. One finds values from 60\,km reported by Kneer et al. (\cite{kneer}), 60\,-\,90\,km (Espagnet et al. \cite{espagnet95}) 170\,km (Hanslmeier et al. \cite{hanslmeier93}; Komm et al. \cite{komm91a}) up to 270\,km where Bendlin \& Volkmer (\cite{bendlin}) detected brightness signatures of the granulation. The values may differ from each other due to differences in the method used to establish the geometrical height scale (transformation from $\tau$ to $z$) and in different methods to filter the oscillations. The vertical velocity field persists in the upper layers, but waves (acoustic and gravity) and turbulent motions may also contribute to the velocity fluctuations (Deubner \cite{deubner88}; Deubner \& Fleck \cite{deubner89}; Salucci et al. \cite{salucci}; Espagnet et al. \cite{espagnet95}).

In many of the studies above, based on the analysis of line parameters, specific positions in the spectral lines have been directly associated with geometrical heights by means of the 'formation heights'. This is not a reliable concept since in reality, the information from different atmospheric layers is mixed in the spectral lines. Therefore, it is more appropriate to study the response functions to a physical parameter, which show the variation of the emergent intensity at a given wavelength produced by a differential change in the specific parameter at a given photospheric layer (S\'anchez Almeida et al. \cite{sanchez96}).

In this series of papers a complete study of the vertical structure of the solar photosphere will be performed. After a first attempt with a global correlation, and spatial coherence and phase spectra analyses of different line parameters in the first paper of this series (Puschmann et al. \cite{puschmann03}, hereafter Paper I), the aim of the present work is to gain insight into  the variation of physical quantities characterising granulation (like temperature ($T$), line of sight velocity ($V_{\rm LOS}$), gas pressure ($P_{\rm g}$) and density ($\rho$)), obtained from the inversion of our time series of one-dimensional slit spectra. The present investigation consists of three parts. Firstly, a study of coherence and phase spectra analysis has been carried out, to gain information about the height variation of physical quantities of structures of different size.  Secondly, we aim at the creation and analysis of a model of an average granular cell, showing the granule to intergranular lane stratification of physical quantities at different optical depths and geometrical heights to retrieve information about the variation of physical quantities of different structures. Thirdly, the horizontal variations of physical quantities for the cases of a specific small and a typical large granular cell at different optical depths and geometrical heights are presented.

\section{Observation}
The data used for this study consist of a 50 minute time series of high resolution quiet-granulation CCD spectrograms, containing three photospheric \ion{Fe}{i} lines ($\lambda$ 6494.98, 6495.74, 6496.47 \AA, hereafter referred to as Line I, Line II, and Line III, respectively). Line II has been excluded from this analysis due to a blend of a terrestrial water vapour line.  Blends in the left wing of Line I and in the right wing of Line III have been considered in the spectral synthesis, hereafter referred to as Line IV and V, respectively (see Table \ref{tab1}).

\begin{table*}[]
\caption[]{Atomic parameters used for the inversion of the spectroscopic time series. The columns denote in order: Line number, element, wavelength (\ion{Fe}{i} lines: Nave et al. \cite{nave}; \ion{Ba}{ii} line: Pierce \& Breckinridge \cite{pierce}), damping enhancement factor, excitation potential (Nave et al. \cite{nave}), oscillator strength (evaluated in this paper, see text for details), quantum numbers of the lower and upper levels of the transition (Nave et al. \cite{nave}), velocity exponent ($\alpha$) and line broadening cross-section ($\sigma_{\rm c}$) of the collisional broadening kindly provided by L.R. Bellot Rubio. Line II has been excluded due to a strong blend with a water vapour line.}
\normalsize
\begin{flushleft}
\begin{tabular}{lllllllll}
\hline 
Line & Element &  $\lambda $ (\AA) & $E$ & $P_{\rm exc}$ (eV)& $\log gf$ &
Transition & $\alpha$ & $\sigma_{\rm c}$(cm$^{2}$)  \\
\hline
I & \ion{Fe}{i} & 6494.9805 & 1.0 & 2.39 & $-$1.29 & 3H 6.0 $-$ 5G 5.0 & 0.247 & 8.98800e$-$15 \\
III & \ion{Fe}{i} & 6496.4666 & 1.0 & 4.77 & $-$0.56 & 3D 2.0 $-$ 3D 2.0 & 0.279 &
2.59000e$-$14 \\
IV & \ion{Fe}{i} & 6494.4999 & 1.0 & 4.71 & $-$1.2 & 3D 3.0 $-$ 5P 2.0 & 0.0 & 0.0 \\
V & \ion{Ba}{ii} & 6496.9095 & 0.1 & 0.60 & \,\,\, 0.1 & 2D 1.5 $-$ 2P 0.5 & 0.0 & 0.0 \\
\hline
\end{tabular}
\end{flushleft}
\label{tab1}
\end{table*}

The data were obtained on July 8, 1993, with the 70 cm Vacuum Tower Telescope at the Observatorio del Teide (Tenerife). The observations have been performed at the solar disk centre. The entrance slit of the spectrograph was set to 100 $\mu$. The pixel size in the focal plane of the spectrograph was 0\farcs093 in the spatial direction and 2.06\,m\AA\, in the spectral direction. In the spatial direction, the slit covered a total length of $95\farcs2$ (1024\,pix) although after flat-fielding, only the central part between the two reference hairs for positioning (46\farcs6 equivalent to 501\,pix) was kept for further analyses. From the computed power spectra we estimate the effective spatial resolution achieved as $\sim$ 0\farcs5.

For further details about observation and data reduction  see Paper I.

\section{Inversion of spectroscopic data}
\label{sir}
The inversion technique SIR (Stokes Inversion based on Response function, Ruiz Cobo \& del Toro Iniesta \cite{ruizcobo92}) provides the stratification of thermodynamic and dynamical parameters like temperature ($T$), line of sight velocity ($V_{\rm LOS}$) and micro-turbulent velocity ($\xi_{\rm mic}$) {\it vs.} continuum optical depth at 5000 \AA\,. Hereafter we refer our optical depths $\tau$ to the standard wavelength $\lambda$\,=\,5000 \AA\, ($\tau\,=\,\tau_{5000}$). The geometrical height scale ($z$), electron pressure ($P_{\rm e}$), gas pressure ($P_{\rm g}$), and density ($\rho$) are also evaluated under the assumption of hydrostatic equilibrium ($\rm HE$) in each iteration but they depend strongly on the boundary conditions. Since a perturbation of these parameters does not significantly affect intensity profiles, the boundary condition can be changed afterwards.

We refer the reader to Ruiz Cobo \& del Toro Iniesta (\cite{ruizcobo92}) for a detailed description of the inversion method. In summary, the technique is a Marquardt non-linear least-square method (see Press et al. \cite{press}), in which the so-called response functions of intensity are used as derivatives of the  $\chi^2$-merit function. The procedure starts with the estimation of a model atmosphere which is modified iteratively until the synthetic profiles, calculated with the new model atmosphere, match the observed ones. The difficulty of the inversion of the radiative transfer equation is precisely its non linearity (in contrast to other inversions like in helio-seismology, which allow the evaluation of the kernels once and they are kept constant throughout the inversion process). The response functions depend so strongly on the model that they have to be calculated in each iteration. Furthermore the iterative process has to be treated carefully to ensure convergence. Ruiz Cobo \& del Toro Iniesta (\cite{ruizcobo92}) demonstrated the convergence of the code to reasonable and unique models, even in case of simulated observations with very poor signal to noise ratios.

The basic assumptions of SIR are local thermo-dynamical equilibrium ($\rm LTE$), a homogeneous plane-parallel atmosphere and hydrostatic equilibrium. The required absorption coefficients, Planck function, damping and Voigt function are calculated by using Wittmann's (\cite{wittmann}) routines modified so that the whole set of partial derivatives is also obtained. 

\subsection{Atomic line parameters}
The reliability of model atmospheres determined from inversions depends strongly on the accuracy of the atomic parameters available for each line. For the inversion of our data set the values presented in Table \ref{tab1} have been considered and are discussed below.

Collisional broadening by neutral hydrogen atoms is known to be underestimated in spectral syntheses using the van der Waals interaction potential due to its asymptotic nature. In order to compensate for the smaller broadening, the micro-turbulence needs to be overestimated. The introduction of the so-called damping enhancement factor ($E$) in order to increase the broadening by collisions has the objective of mitigating this problem. However because of the ad-hoc nature of this correction factor, the micro-turbulence from observed spectral lines is expected to be questionable. Fortunately, the use of $E$ is no longer necessary following the important advances achieved on the quantum formulation of the collisional broadening (O'Mara \cite{omara76}; Anstee \& O'Mara \cite{ansteeomara95}, Barklem et al. \cite{barklemomara00}, and references therein), which make a realistic estimation of the interaction potential with neutral perturber atoms.  Therefore for the inversion we have used a new version of SIR (Borrero \& Bellot Rubio \cite{borrero}) implementing this new treatment of collisional broadening. The necessary parameters $\alpha$ (velocity exponent) and $\sigma_{\rm c}$ (line broadening cross-section) needed to calculate the collisional broadening of our lines have been kindly provided by L.R. Bellot Rubio (see Table \ref{tab1}).

Abundances have been taken from Th\'evenin (\cite{thevenin}) and the quantum numbers of the lower and upper levels of the transition, central laboratory wavelengths and the excitation potential ($P_{\rm exc}$) from Nave et al. (\cite{nave}). 

\begin{figure}[]
\centering
\includegraphics[width=8cm]{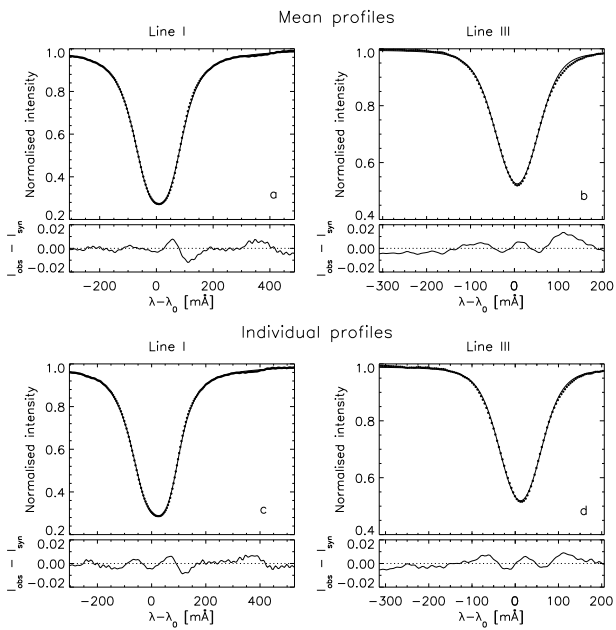}
\caption []{Plot of the observed and synthesised profiles using the atomic parameters given in Table \ref{tab1}. Upper panels: Mean observed profiles (dotted) and synthesised profiles (solid) and the difference (small panels) between them. Lower panels: Same but for an individual case.}
\label{fig1}
\end{figure}

The adequate values of $\log gf$ (presented in Table \ref{tab1}) have been derived from the best-fit between observed and synthesised profiles (see Fig. \ref{fig1}). 

\label{inversiondata}
\subsection{Depth grid}
For the inversion, the optical depth grid has been chosen as equi-spaced on a logarithmic scale, including 37 points with a step width of 0.2, spanning from $\log\tau$\,=\,1.2 to $-$\,6.0. The various nodes in which the physical quantities are evaluated by SIR cover the same range of optical depth. The highest and the deepest points of the grid are always included in the set of nodes. The whole iterative process is divided into separated groups of iterations, each characterised by the number of nodes selected. One of these groups runs iteratively until convergence is reached, i.e., until the variation of $\chi^2$ between any two successive iterations is of no significance. The resulting model atmosphere enters, as an initial one, the next group of iterations with a larger number of nodes. By this procedure, it is possible to slowly decrease the number of degrees of freedom. Therefore, intermediate models are smooth. For the inversion of our data, 3 iterations have been chosen, fixing  the number of nodes for $T$ to 2, 5, 7 and for $V_{\rm LOS}$ to 2, 4, 6 respectively. Electron pressure $P_{\rm e}$, gas pressure $P_{\rm g}$, and geometrical height $z$ are derived from the $T$ stratification at each iteration assuming hydrostatic equilibrium (HE). Micro-turbulence has been assumed to be constant with depth. Additionally a macro-turbulence velocity has been derived for each pixel.

\subsection{$\rm NLTE$--effects}
SIR calculates the atomic level populations under the $\rm LTE$ approximation. Nevertheless, departures from $\rm LTE$ have proved to be important for Line I (see Paper I). A detailed investigation of these effects has recently been presented by Shchukina \& Trujillo Bueno (\cite{Shchukina}). The departures affect mainly the line source function, which deviates significantly from Planck's function as one proceeds outwards in the atmosphere. We have used departure coefficients $\beta_{\rm low}\,(\tau)$ and $\beta_{\rm up}\,(\tau)$ obtained by Shchukina \& Trujillo Bueno for the quiet sun model of Maltby et al. (\cite{maltby} MACKKL model) and a 63 level model atom. $\beta_{\rm low}$ ($\beta_{\rm up}$) stands for the ratio between the population of the lower (upper) atomic level evaluated in $\rm NLTE$ and $\rm LTE$ conditions, respectively. 

$\rm NLTE$ effects have to be considered for both synthesis of the line
profiles and calculation of the response function. The results will be
more reliable, the more similar the final model (obtained by the inversion)
and the MACKKL model are. Deviations of the final model from the
MACKKL model in deep layers do not produce significant errors, because the
departure coefficients at those layers are always near unity for solar
models. However, in high layers deviations of the final model from the
MACKKL model could produce significant changes in the departure coefficients (see Paper I).

\subsection{Resulting stratifications of physical quantities as functions of optical depth and geometrical height}
\label{transf}
The 3D box of intensity $I(x_{\rm i},\,t_{\rm j},\,\lambda_{\rm k})$, where $i$\,=\,1, 500 with steps of 0\farcs093, $j$\,=\,1, 152 with steps of 20 sec, $k$\,=\,1, 1024 with steps of 2.06 m\AA\, has been inverted and therefore for each spatial and temporal point the stratifications of the physical quantities  $T$, $V_{\rm LOS}$, $P_{\rm g}$, $z$ and $\rho$ as functions of the continuum optical depth at 5000 \AA\, ($\tau_{5000}$) have been obtained. The uncertainties of the derived physical quantities are calculated from the difference between the observed and synthetic spectra, following the standard non-linear least squares error propagation. Due to large error bars at optical depths smaller than $\log\tau$\,=\,$-$\,4, only the optical depth range between 1.2 and $-$\,4 is used for further computations. Since the observations are obtained at disc centre, $V_{\rm LOS}$ will be termed hereafter $V_{\rm z}$ --  vertical velocity.

Subsequently the 3D boxes, containing the spatial and temporal variation of physical quantities {\it vs.} $\log\tau$ have been filtered layer by layer of oscillations by applying a subsonic filter in the Fourier space ($k_{\rm x},\omega$), described in Paper I. We have chosen the maximum phase velocity admitted by the filter being 5\,km\,s$^{-1}$ and a gradual cutoff of a cosine bell between the lines $\omega=v_p\,k_x$  and  $\omega=1.08\,v_p\,k_x$ has been used.

\begin{figure}[]
\centering
\includegraphics[width=8.5cm]{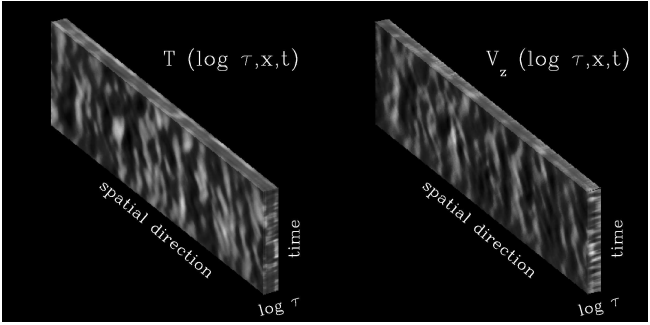}
\caption []{Resulting 3D boxes of the physical quantities $T$ and $V_{\rm z}$ as functions of $\log\tau_{\rm 5000},\,x,\,t$ after filtering oscillations. $T$ and $V_{\rm z}$ are represented, with $\log\tau$ running from 0 to $-$\,4 (step width of 0.2), $x$ running from 1 to 500 (step width of 0\farcs093) and $t$ running from 1 to 152 (step width of 20 sec).}
\label{fig2}
\end{figure}

\noindent
As an example, in Fig. \ref{fig2} the filtered 3D boxes of $T\,(\log\tau,\,x,\,t)$ and $V_{\rm z}\,(\log\tau,\,x,\,t)$ are presented.  Note that at $\log\tau$\,=\,0 the structures are almost identical to those of the intensities presented in Paper I. 

The geometrical scale resulting from the inversion of the data is different for each spatial and temporal point. The criterion to establish a common height scale has been to neglect horizontal acceleration at one specific height, that is, imposing horizontally constant gas pressure at this layer. A constant gas pressure value of $P_{\rm g}^{\rm M}$\,=\,1.82\,$\times$\,$10^{5}$\,dyn\,cm$^{-3}$ (MACKKL model) at a height of $z_{\rm 0}$\,=\,$-$\,80\,km ($\log\tau$\,=\,1.08) has been chosen. Subsequently the needed shift $\delta z\,(x,\,t)$ has been evaluated so that 
\begin{equation}
P_{\rm g}(z=z_{\rm 0}-\delta z)=P_{\rm g}^{\rm M}.
\end{equation}
Finally, each physical quantity $T\,(z,\,x,\,t)$, $V_{\rm LOS}\,(z,\,x,\,t)$, $P_{\rm g}\,(z,\,x,\,t)$, $\rho\,(z,\,x,\,t)$ has been interpolated to a new equi-spatial grid within a $z$ range from $-$\,120\,km to 400\,km, with a step width of 20\,km. The geometrical level $z$\,=\,0 corresponds to $\tau_{5000}$\,=\,1.

\section{Spatial coherence and phase spectra analysis}
\label{cohphase}
An analysis by means of spatial coherence and phase difference spectra (see Paper I for a detailed description of this method) between temperature ($T$) and vertical velocity ($V_{\rm z}$) fluctuations at the consecutive levels throughout the solar photosphere has been used to study the height variation of physical quantities of structures of different size.

\subsection{Physical quantities in optical depth}
\label{phasecohtau}

\begin{figure}[]
\centering
\includegraphics[width=8.6cm]{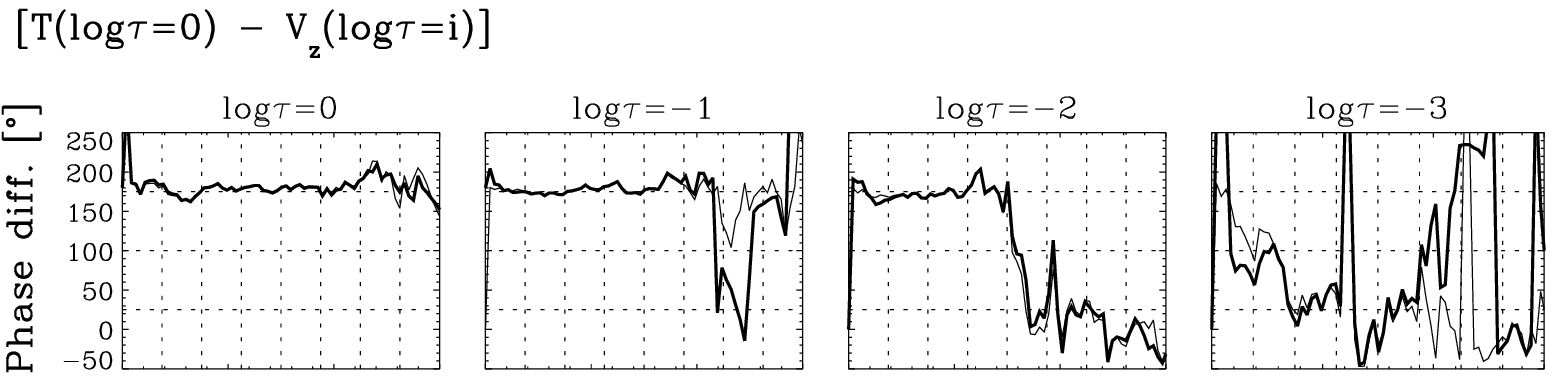}
\includegraphics[width=8.6cm]{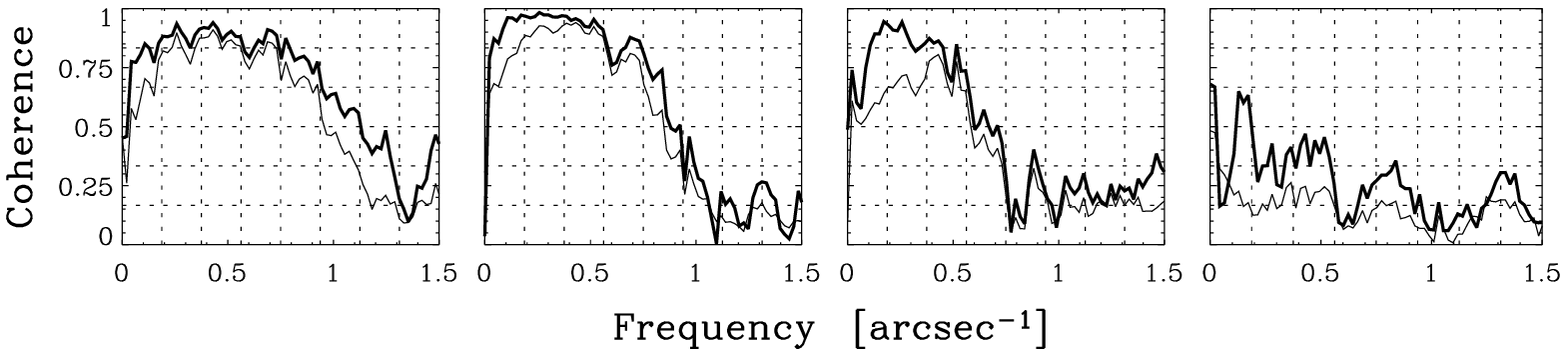}
\includegraphics[width=8.6cm]{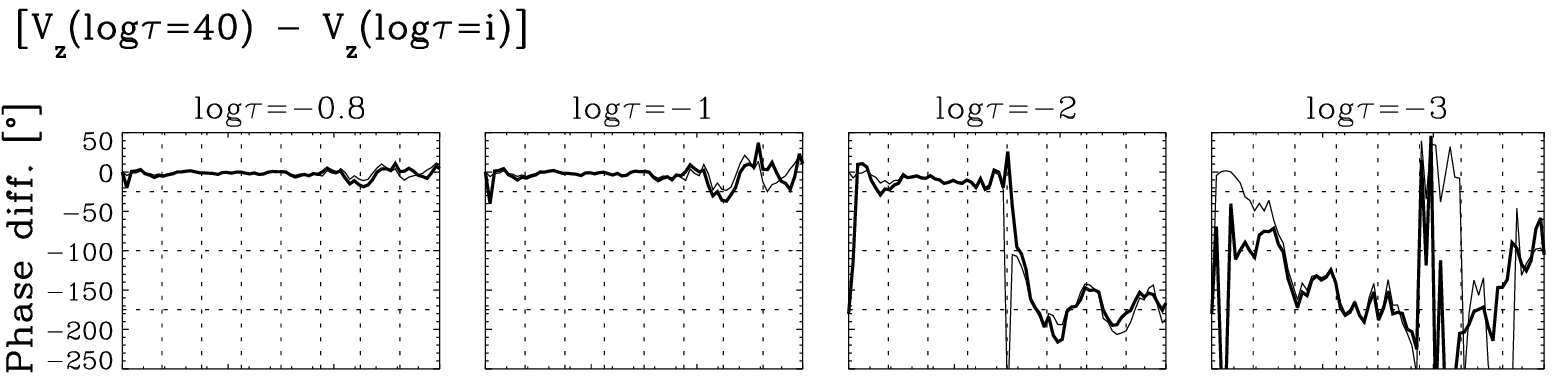}
\includegraphics[width=8.6cm]{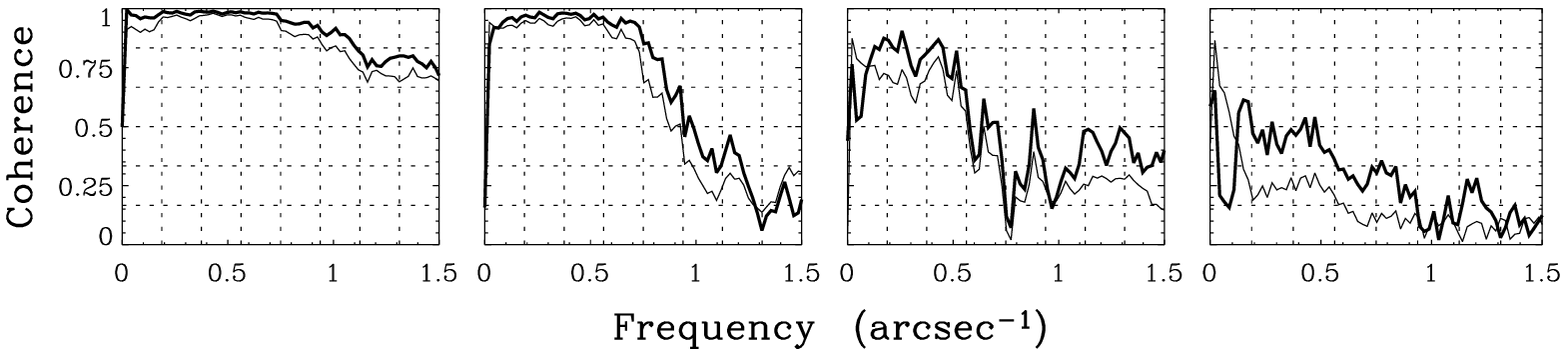}
\includegraphics[width=8.6cm]{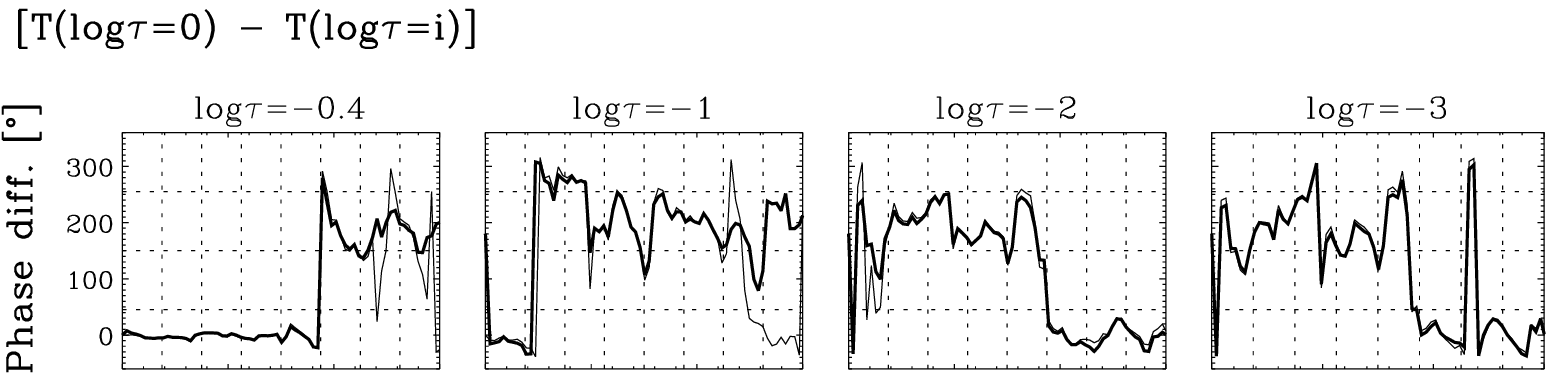}
\includegraphics[width=8.6cm]{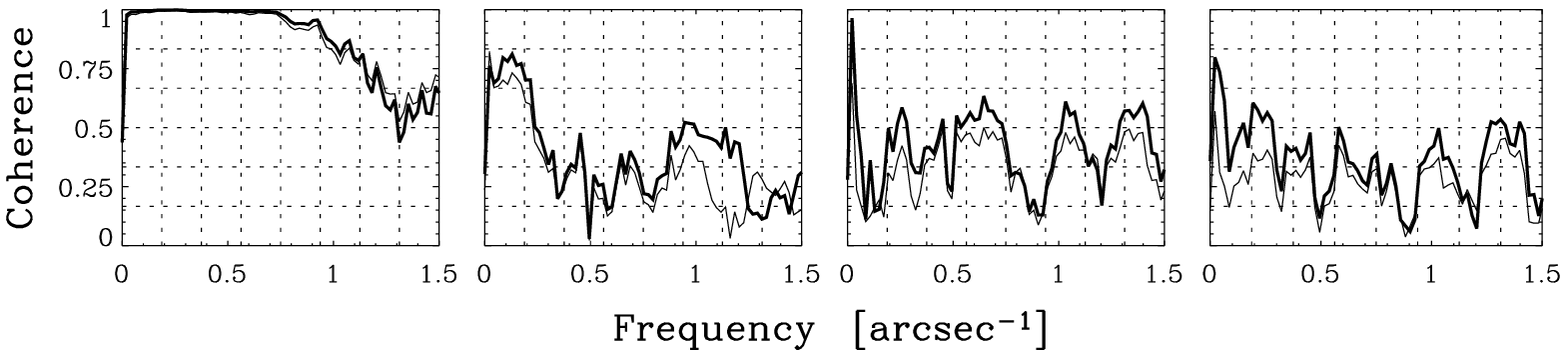}
\caption []{Phase difference and coherence  {vs.} spatial frequency for unfiltered (thin lines) and filtered data (thick lines) between the pairs: [$T\,(\log\tau$\,=\,$0$)\,--\,$V_{\rm z} \,(\log\tau$\,=\,$i)$], where $i$\,=\,$0,-1,-2,-3$; [$V_{\rm z}\,(\log\tau$\,=\,$-0.4)$\,--\,$V_{\rm z} \,(\log\tau$\,=\,$i)$] and [$T\,(\log\tau$\,=\,$0$\,)\,--\,$T\,(\log\tau$\,=\,$i)$], where $i$\,=\,$-0.8,-1,-2,-3$ and  $i$\,=\,$-0.4,-1,-2,-3$, respectively.}
\label{cohphatau}
\end{figure}

Coherence and phase shifts {\it vs.} spatial frequency  $u$ between temperature and vertical velocity (i.e., between the pairs [$T\,(\log\tau$\,=\,$0$)\,--\,$V_{\rm z} \,(\log\tau$\,=\,$i)$] for $i$\,=\,$0,-1,-2,-3$, [$V_{\rm z}\,(\log\tau$\,=\,$-0.4)$\,--\,$V_{\rm z} \,(\log\tau$\,=\,$i)$] for $i$\,=\,$-0.8,-1,-2,-3$, and [$T\,(\log\tau$\,=\,$0$\,)\,--\,$T\,(\log\tau$\,=\,$i)$] for $i$\,=\,$-0.4,-1,-2,-3$) resulting from temporal average, are presented in Fig. \ref{cohphatau} for unfiltered and filtered data, respectively.

\begin{figure*}[]
\centering
\includegraphics[width=7.7cm]{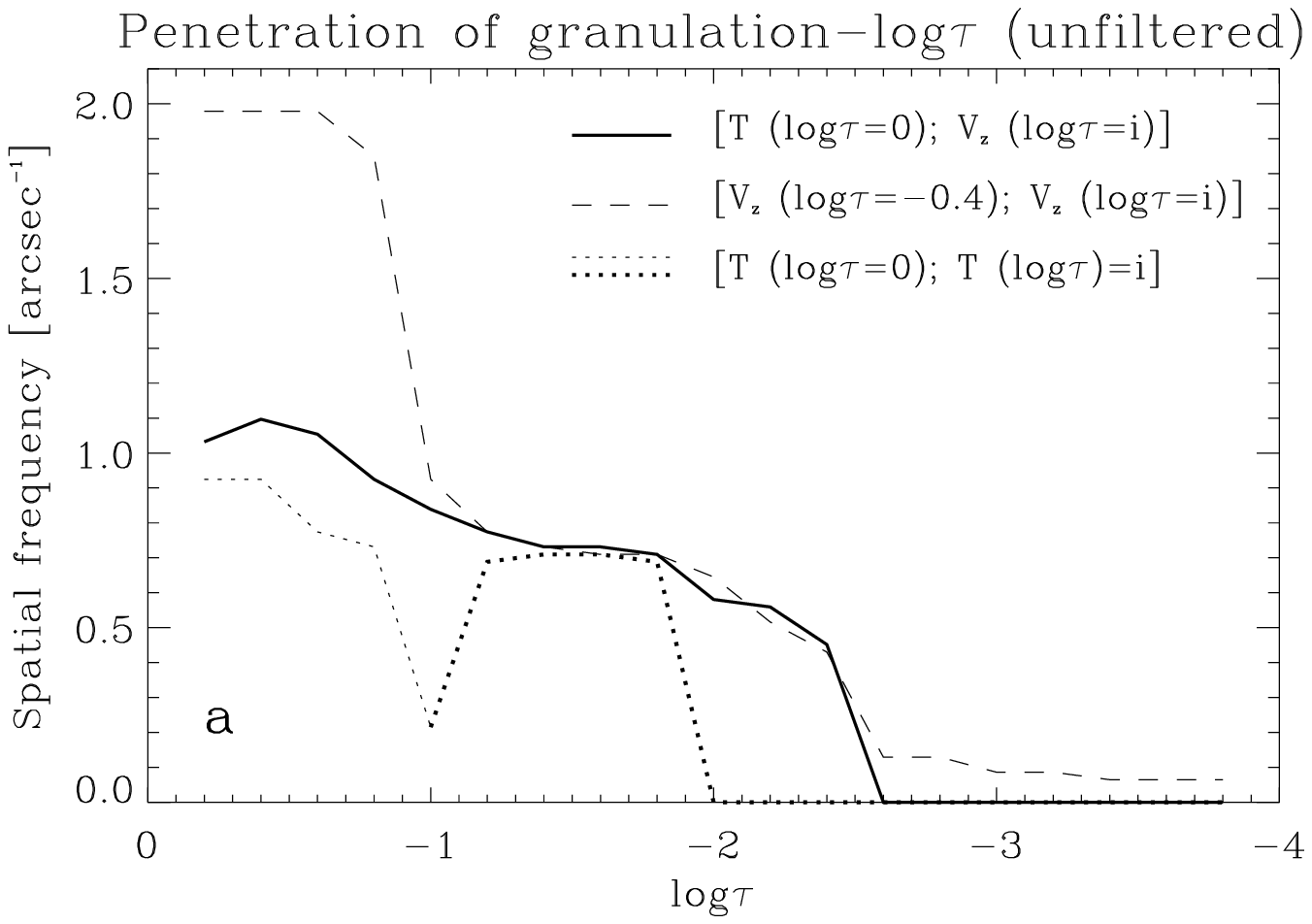}
\hspace{-1.8cm}
\includegraphics[width=7.7cm]{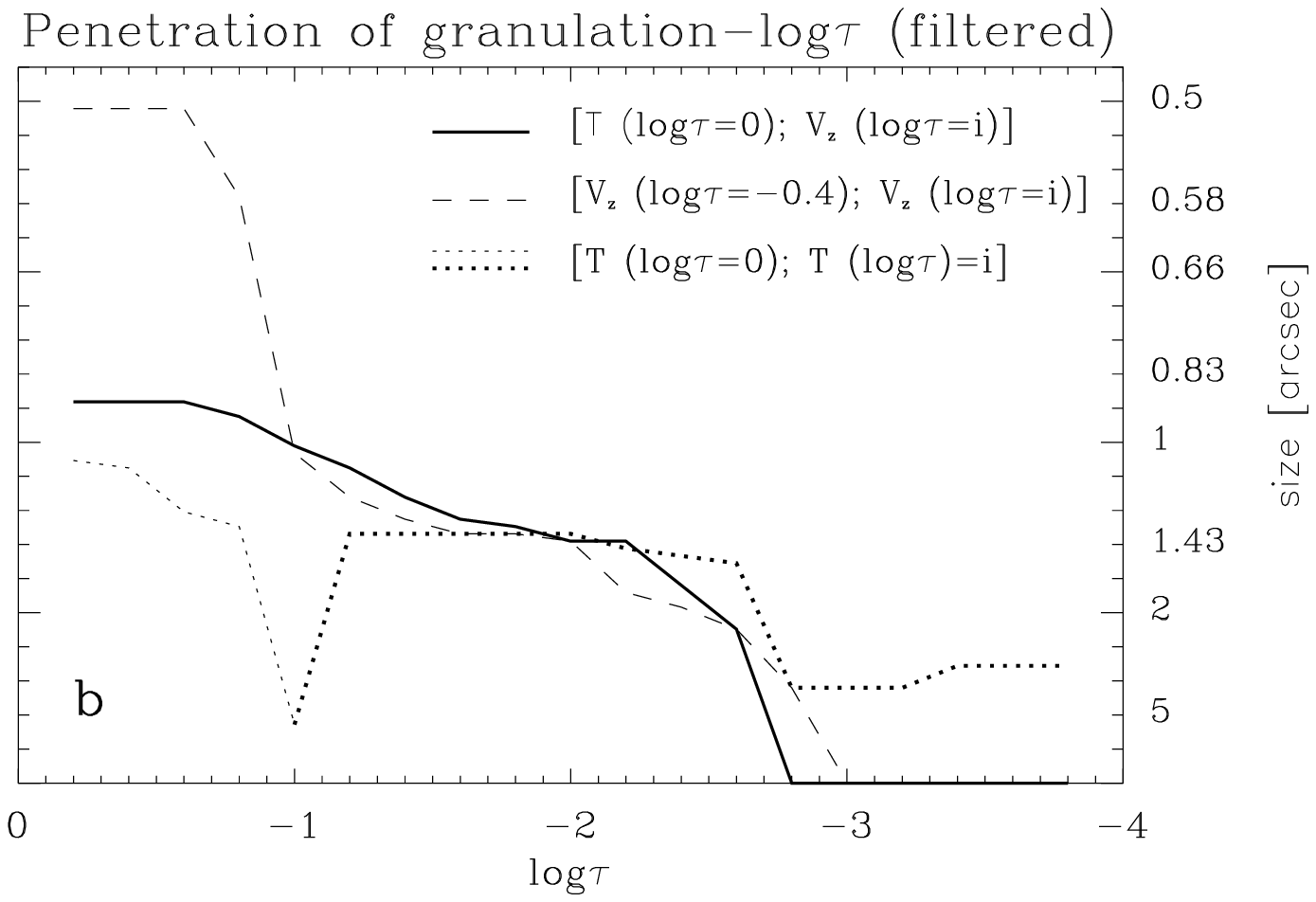}
\caption []{Maximum (minimum) spatial frequencies (sizes) of structures contributing to the $T$ and $V_{\rm z}$ fluctuations at different optical depths, as obtained from spatial coherence and phase difference analyses  {\it vs.} spatial frequency between the pairs [$T\,(\log\tau$\,=\,$0$)\,--\,$V_{\rm z}\,(\log\tau$\,=\,$i)$], [$V_{\rm z}\,(\log\tau$\,=\,$-0.4$)\,--\,$V_{\rm z}\,(\log\tau$\,=\,$i)$] and  [$T\,(\log\tau$\,=\,$0$)\,--\,$T\,(\log\tau$\,=\,$i)$], $i$\,=\,$-0.2,\dots,-4$, $\Delta i$\,=\,$0.2$, for unfiltered (panel a) and filtered data (panel b), respectively. See the text for a detailed description.}
\label{hdeptau}
\end{figure*}

To facilitate the estimation of the different penetration heights  of $T$ and $V_{\rm z}$ depending on the structure size, the maximum spatial frequency (minimal size) of structures contributing to the temperature and velocity fluctuations at each optical depth are presented in Fig. \ref{hdeptau}. This figure has been constructed  by assuming the following constraint: We consider only spatial frequencies $u$\,$<$\,2 arcsec$^{-1}$ (structures\,$>$\,0\farcs5), a coherence\,$>$\,0.5, and a phase difference of  $180^{\circ}\,\pm$\,20$^{\circ}$ between temperature and vertical velocity at the different optical depths $i$, whereas between the velocities a $0^{\circ}\,\pm$\,20$^{\circ}$ phase difference is assumed. Between temperatures a $0^{\circ}\,\pm$\,20$^{\circ}$ phase difference at layers below $\log\tau$\,=\,$-$1 (thin dotted line of Fig. \ref{hdeptau}) and a phase difference of $180^{\circ}\,\pm$\,20$^{\circ}$ at layers above (thick dotted line of Fig. \ref{hdeptau}) is considered. As the  lower and upper limit of granular scales we assume $\sim$\,0\farcs5, as the expected limit of the spatial resolution of our data set, and $\sim$\,4\arcsec, respectively. 

For filtered data in general a larger penetration height for all physical quantities is found compared with unfiltered data. From the coherence and phase shift analysis between the pair [$T\,(\log\tau$\,=\,$0$)\,--\,$T\,(\log\tau$\,=\,$i)$] at lower layers a $0^{\circ}$ phase difference together with a coherence\,$>$\,0.5 at granular scales is detected (see Fig. \ref{cohphatau} and thin dotted line of Fig. \ref{hdeptau}). A fast decay of temperature fluctuations with decreasing optical depth together with a decreasing contribution of small structures is revealed, vanishing at $\log\tau$\,=\,$-$1. At higher layers and at granular scales $>$\,1\farcs5  an inversion of the temperature fluctuations is observed, reflected in the phase jump from  0$^{\circ}$ to 180$^{\circ}$  together with a coherence\,$>$\,0.5 (see Fig. \ref{cohphatau} and thick dotted line of Fig. \ref{hdeptau}). These findings confirm the results obtained after an analysis of global correlations, spatial coherence and phase spectra between the fluctuations of different line parameters presented in Paper I.

For vertical velocities, the coherence and phase shifts between the pairs [$T\,(\log\tau$\,=\,$0$)\,--\,$V_{\rm z} \,(\log\tau$\,=\,$i)$] and [$V_{\rm z}\,(\log\tau$\,=\,$40$)\,--\,$V_{\rm z} \,(\log\tau$\,=\,$i)$] reveal a similar size distribution of velocity and temperature structures with height, as expected from the overshoot scenario. Vertical velocities penetrate into the highest layers under study, even at $\log\tau$\,=\,$-$2.8 convective velocities of the larger granular structures ($\sim$\,4\arcsec) are present. A less efficient height penetration of smaller structures is found, although still at $\log\tau$\,$\sim$\,$-$2 structures $>$\,1\farcs4 are detected.

\subsection{Physical quantities in geometrical height}
\label{phasecohgeo}

Properties observed at equal optical depths (e.g. $\log\tau$\,=\,0) do not refer to a unique geometric depth and observers should be cautious in their interpretations about temperatures, Doppler velocities, etc. This is a result of the rapid increase in the $H^-$ opacity with temperature, so one looks down to shallower depths in hotter regions and does not see the high-temperature gas, because it is very opaque. As an example, much smaller temperature fluctuations are observed at equal optical depths than at equal geometrical levels (see Figs. \ref{clust1} and \ref{clust2}). Keeping this in mind, the presentation and analysis of physical quantities at different geometrical heights facilitates the interpretation of the results.

\begin{figure}[]
\centering
\includegraphics[width=8.7cm]{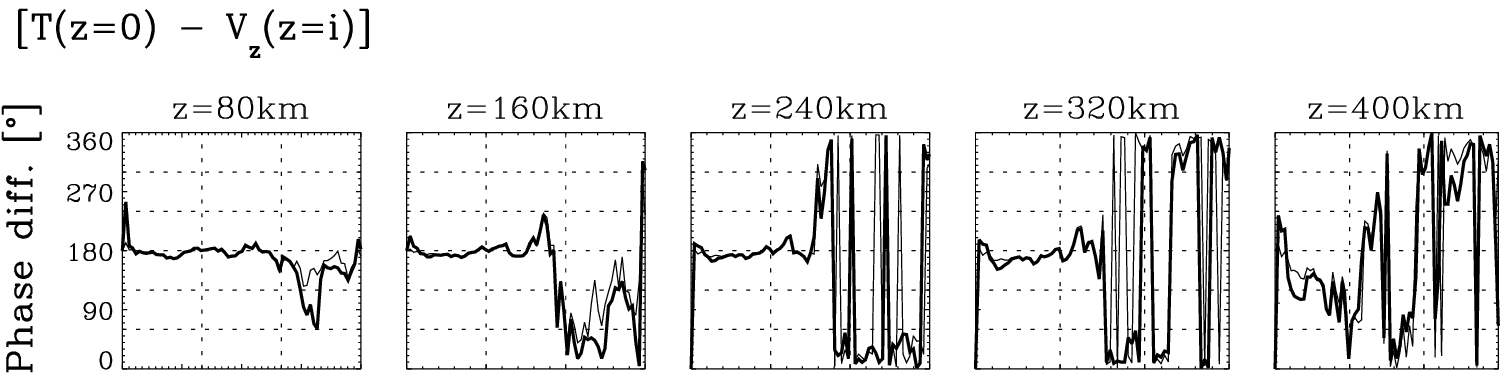}
\includegraphics[width=8.7cm]{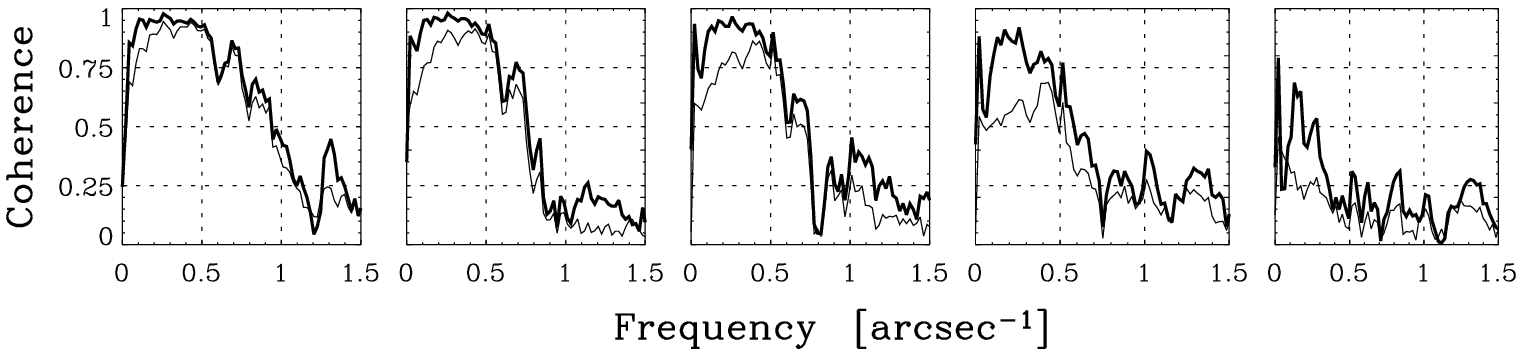}
\includegraphics[width=8.7cm]{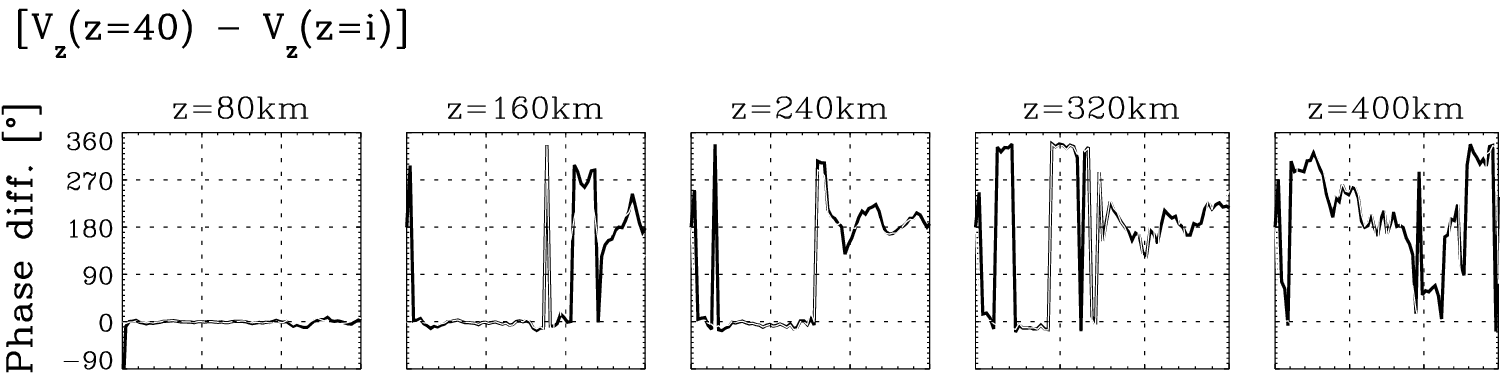}
\includegraphics[width=8.7cm]{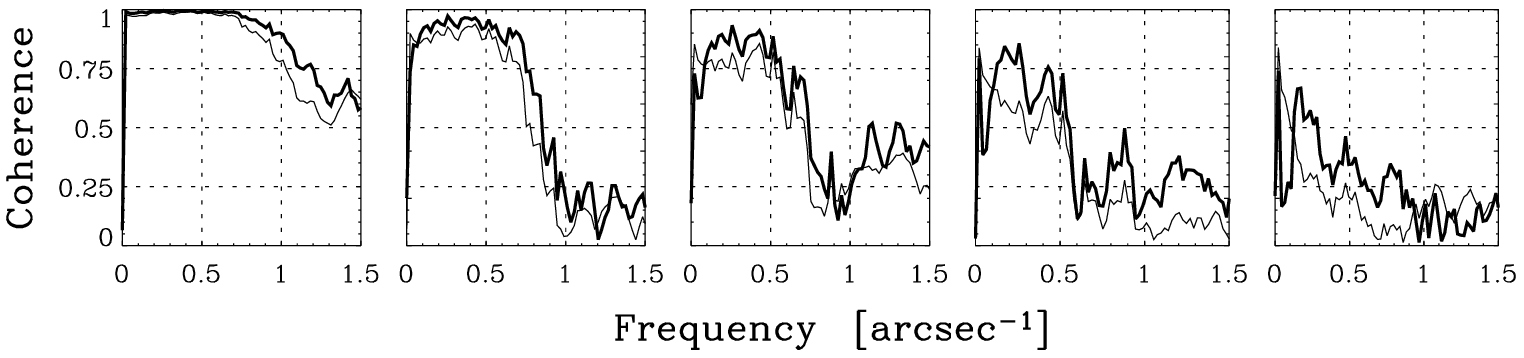}
\includegraphics[width=8.7cm]{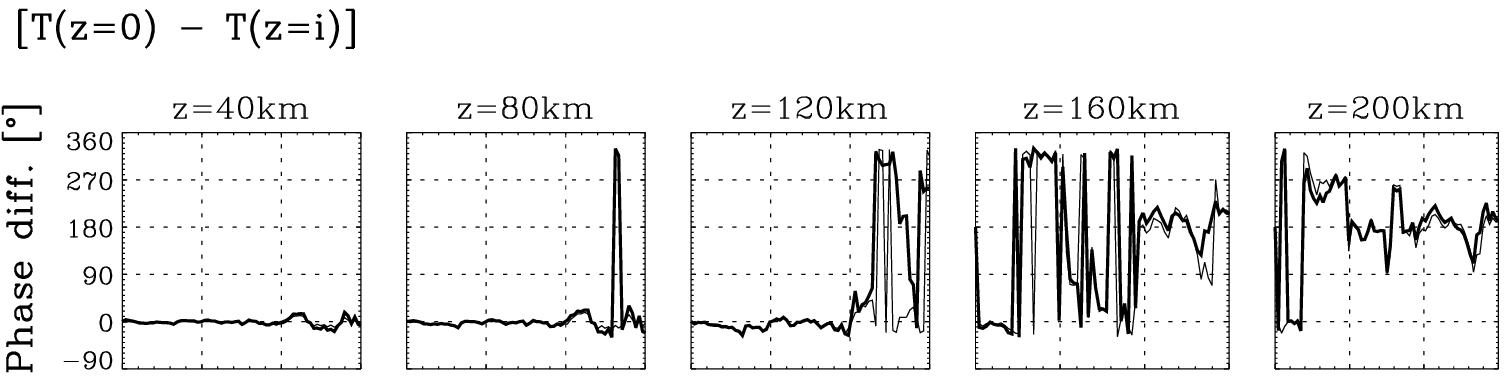}
\includegraphics[width=8.7cm]{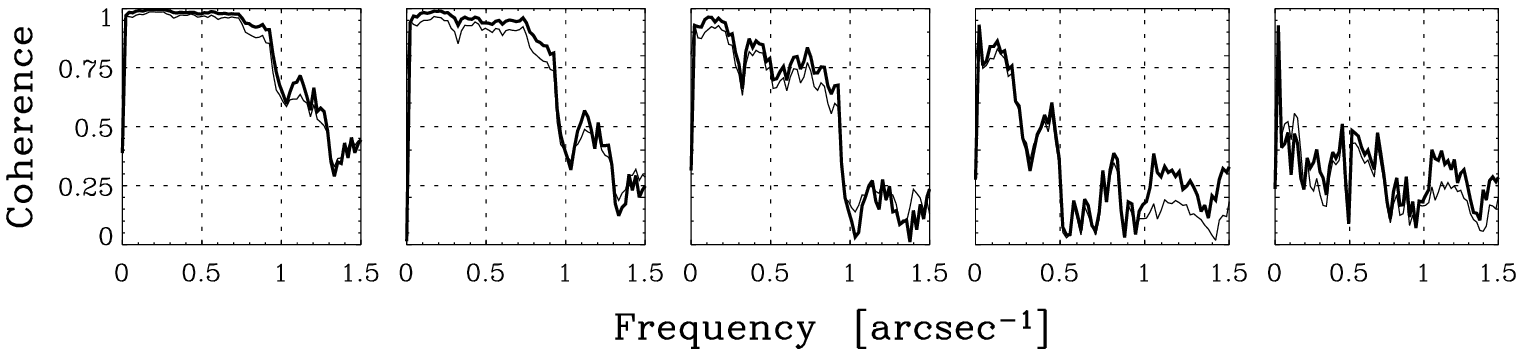}
\caption []{Phase difference and coherence  {vs.} spatial frequency for unfiltered (thin lines) and filtered data (thick lines) between the pairs: [$T\,(z$\,=\,$0$\,km)\,--\,$V_{\rm z} \,(z$\,=\,$i)$] and [$V_{\rm z}\,(z$\,=\,$40$\,km)\,--\,$V_{\rm z} \,(z$\,=\,$i)$], where $i$\,=\,$80,\dots,400$\,km, $\Delta i$\,=\,$80$\,km; [$T\,(z$\,=\,$0$\,km)\,--\,$T\,(z$\,=\,$i)$], where $i$\,=\,$40,\dots,200$\,km, $\Delta i$\,=\,$40$\,km.}
\label{cohpha}
\end{figure}

Coherence and phase shifts {\it vs.} spatial frequency between temperature and vertical velocity (i.e. between the pairs [$T\,(z$\,$=$\,$0$\,km)\,--\,$V_{\rm z}\,(z$\,$=$\,$i)$], [$V_{\rm z}\,(z$\,$=$\,$40$\,km)\,--\,$V_{\rm z}\,(z$\,$=$\,$i)$], [$T\,(z$\,$=$\,$0$\,km)\,--\,$T\,(z$\,$=$\,$i)$], resulting from temporal averages, are presented in Fig. \ref{cohpha}, for unfiltered and filtered data, respectively. $i$ denotes the consecutive layers in geometrical height throughout the solar photosphere. 

\begin{figure*}[]
\centering
\includegraphics[width=7.7cm]{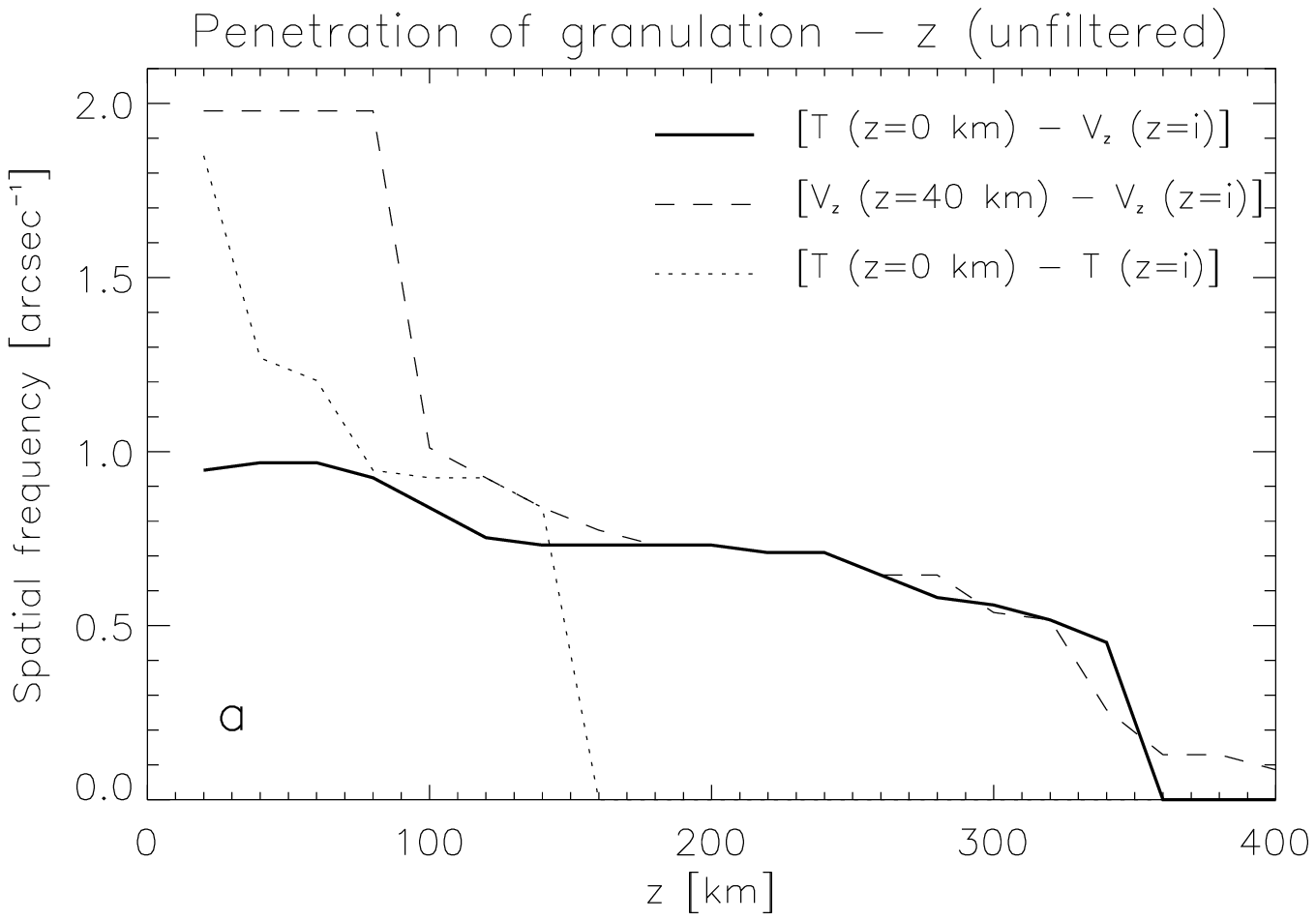}
\hspace{-1.8cm}
\includegraphics[width=7.7cm]{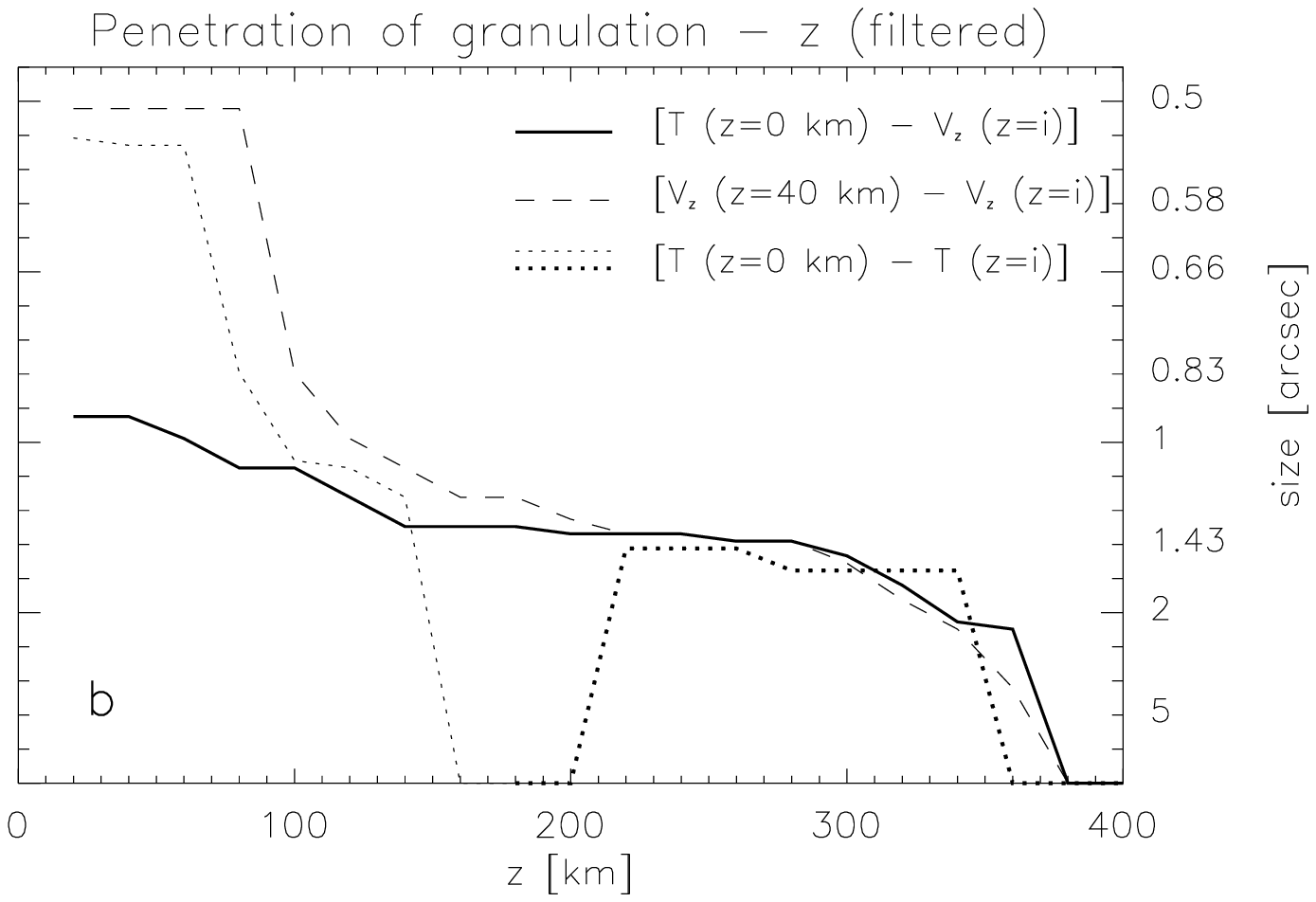}
\caption []{Maximum (minimum) spatial frequencies (sizes) of structures contributing to the $T$ and $V_{\rm z}$ fluctuations at different geometrical heights, as obtained from spatial coherence and phase difference analyses  {\it vs.} spatial frequency between the pairs [$T\,(z$\,=\,$0$\,km)\,--\,$V_{\rm z}\,(z$\,=\,$i)$], [$V_{\rm z}\,(z$\,=\,$40$\,km)\,--\,$V_{\rm z}\,(z$\,=\,$i)$] and  [$T\,(z$\,=\,$0$\,km)\,--\,$T\,(z$\,=\,$i)$], $i$\,=\,$20,\dots,400$\,km, $\Delta i$\,=\,$20$\,km, for unfiltered (panel a) and filtered data (panel b), respectively. See the text for a detailed description.}
\label{hdep}
\end{figure*}

The maximum spatial frequency (minimal size) of structures contributing to the temperature and velocity fluctuations at each geometrical height level are presented in  Fig. \ref{hdep}. This figure has been constructed like Fig. \ref{hdeptau}, described in the previous section. Between temperatures a $0^{\circ}\,\pm$\,20$^{\circ}$ phase difference at layers below 180\,km (see Fig. \ref{hdep}, thin dotted line) and a phase difference of $180^{\circ}\,\pm$\,20$^{\circ}$ at layers above (see Fig. \ref{hdep}, thick dotted line) has been considered. As for optical depth, as the  lower and upper limit of granular scales we have assumed $\sim$\,0\farcs5 (as the expected limit of the spatial resolution of our data set) and $\sim$\,4\arcsec, respectively.

Again for filtered data in general a larger penetration height for all physical quantities is found compared with unfiltered data. From the coherence and phase shift analysis between the pairs [$T\,(z$\,=\,$0$\,km)\,--\,$T\,(z$\,=\,$i)$]  the fast decay of temperature fluctuations together with a decreasing contribution of small structures with  height is revealed. At the layers below 150\,km and at granular scales, we find a coherence larger than 0.5 together with a phase shift of 180$^{\circ}$ between temperature and vertical velocity, whereas temperatures at the different levels are in phase. At layers higher than 200\,km again, as seen in optical depth, an inversion of temperature fluctuations for structures $>$\,1\farcs5 can be observed, but only in case of filtered data, reflected in a phase jump between the pair [$T\,(z$\,=\,$0$\,km)\,--\,$T\,(z$\,=\,$i)$] from  0$^{\circ}$ to 180$^{\circ}$.

For vertical velocities, the coherence and phase shifts between the pairs [$T\,(z$\,=\,$0$\,km)\,--\,$V_{\rm z} \,(z$\,=\,$i)$] and [$V_{\rm z}\,(z$\,=\,$40$\,km)\,--\,$V_{\rm z} \,(z$\,=\,$i)$] confirm the results found for optical depth. The largest granular structures ($\sim$\,4\arcsec) are still observed at a height of 370\,km, whereas structures larger than 1\farcs43 contribute until a height of 280\,km.

Espagnet et al. (\cite{espagnet95}) performed a phase and coherence spectral analysis between intensity and velocity fluctuations as measured at several positions (wavelengths) throughout the profile of the \ion{Na$\rm D_{2}$}{} line ($\lambda$ 5890) which correspond to various heights in the photosphere ranging from the continuum level up to $\sim$ 550\,km. In contrast to our results they conclude that temperature fluctuations of granulation do not penetrate higher than about 60\,-\,90\,km. By comparing the intensity fluctuations at the continuum level with those at different levels in the photosphere, they find high coherence for large granules ($>$1\farcs4) up to 60\,km, vanishing at 90\,km. For Nesis et al. (\cite{nesis88}) and Komm et al. (\cite{komm90}) the coherence of granules $>$ 1\farcs4 vanishes at 170\,km above the continuum level.

There are discrepancies in the literature about the smallest size of velocity structures reaching the 200\,km level above the surface. The values range from 0\farcs9 (Deubner \cite{deubner88}) to 1\farcs5 (Nesis et al. \cite{nesis88}) and up to 2\arcsec of Durrant \& Nesis (\cite{durrant82}). At this height, we obtain significant coherence for structures larger than $\sim$\,1\farcs3. Salucci et al. (\cite{salucci}) and Espagnet et al. (\cite{espagnet95}) claim that the velocity fluctuations remain coherent across the whole photosphere. Espagnet et al. (\cite{espagnet95}) compared the velocities at different layers with both the continuum intensity and the velocity at the height of 30\,km. From these coherence and phase shift analyses they conclude that the overshooting velocities of granules $>$\,1\farcs4 and $>$\,1\farcs6, respectively, cross the whole photosphere up to a level of at least 550\,km. As stated above, we detect only structures $\sim$\,4\arcsec contributing to the velocity field observed at $z$\,$\sim$\,370\,km.

The present work, in contrast to the abovementioned studies found in the literature, is based on the application of an inversion method to estimate the penetration height of granulation. This allows a precise determination of the height where the observed perturbation of physical parameters takes place. Traditional methods, based on contribution functions or on the calculation of bisectors, can introduce large errors in the height determination (S\'anchez Almeida et al. \cite{sanchez96} and references therein).

\section{Model of an average granular cell} 

Taking into account all pixels of our time series, we have 76\,000 points available, where for each of them its model atmosphere is known, i.e. the stratification of $T$, $V_{\rm z}$,  $P_{\rm g}$, $\rho$ {\it vs.} $z$ as well as {\it vs.} $\log\tau$. In order to obtain a model of an average granular cell these model atmospheres have been grouped into 76 bins, where each bin is the average of 1000 individual model atmospheres. As a criterion for this binning we have chosen the temperature values of each bin at the continuum layer $\log\tau$\,=\,0 and $z$\,=\,0, respectively. Thus, the resulting models of our average granular cells, both in $\log\tau$ and $z$, consist of 76 averaged model atmospheres (bins), where the centre of the granule and its surrounding intergranular lane have been assigned to the hottest and coolest bin, respectively, whereas bins with temperature values in between correspond to the gradual transition granule -- intergranular lane. This is an assumption which might not be always fulfilled (i.e. exploding granules), but it seems reasonable to apply this hypothesis for the case of an average of granular cells. Evidently our model is not representative of the case of exploding granules. However exploding granules represent only a small part of granules; according to Mehltretter (\cite{mehltretter78}) their number density is around 4\% and according to Namba (\cite{namba86}) exploding granules cover about 2.5\% of the observed area.

\subsection{Results as a function of optical depth}
\label{modavtau}

\begin{figure}[t]
\centering
\hspace{-0.15cm}
\includegraphics[width=8.5cm,height=8.5cm]{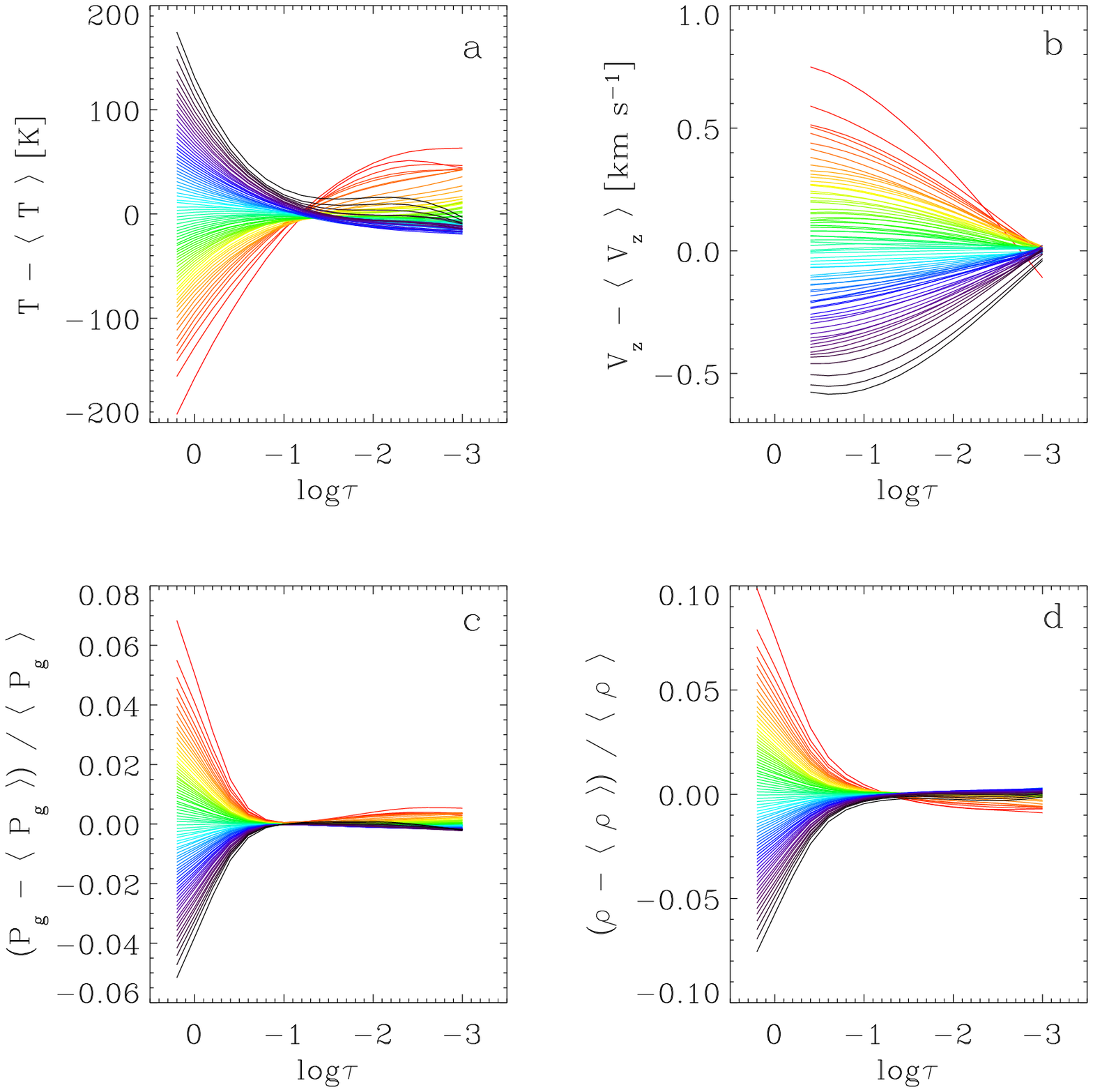}
\includegraphics[width=8.5cm,height=8.5cm]{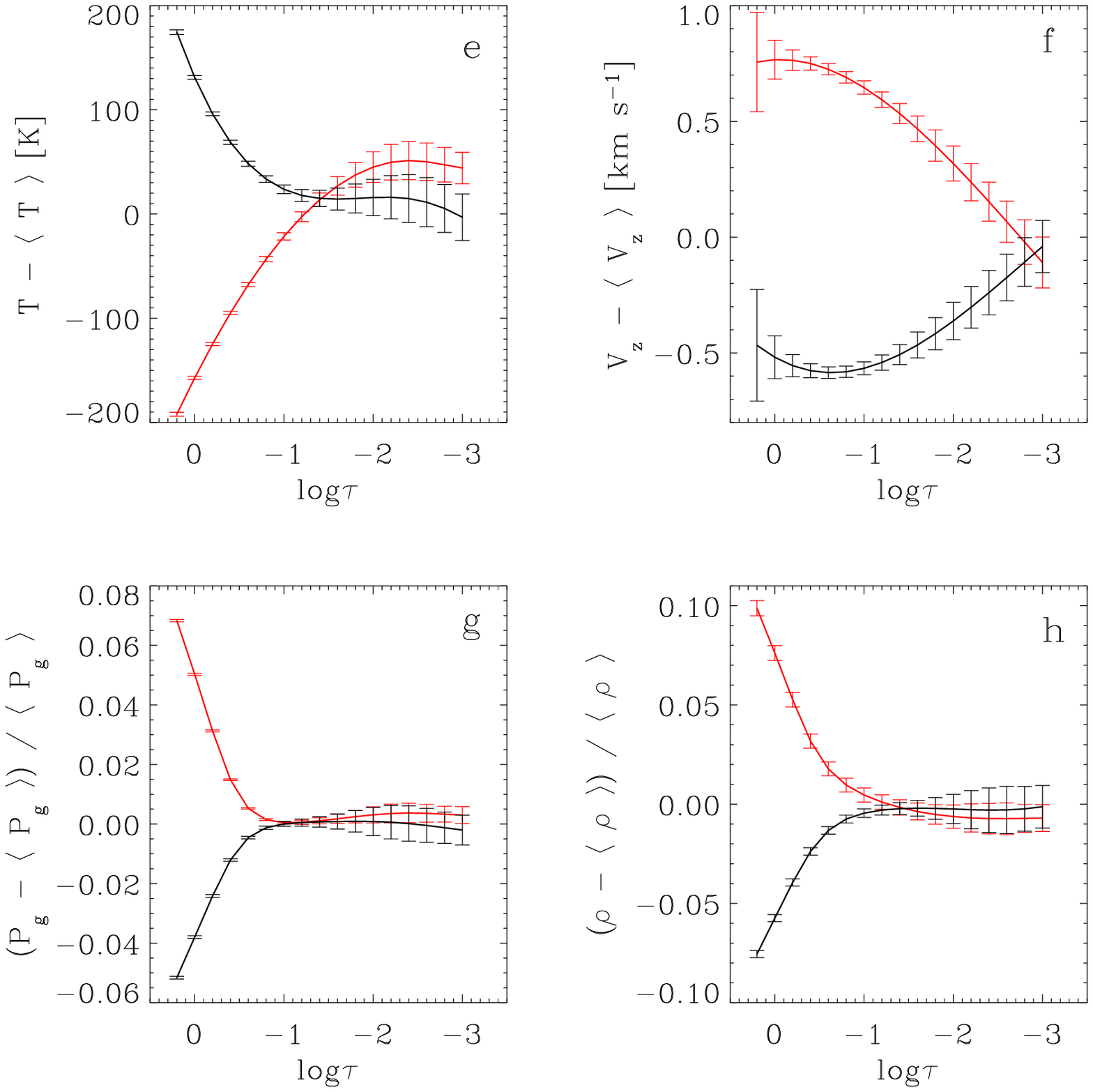}
\caption []{Model of an average granular cell: $T$--$\overline{T}$ (panel a), $V_{\rm z}$--$\overline{V_{\rm z}}$ (panel b), $(P_{\rm g}$\,--\,$\overline{P_{\rm g}})$\,/\,$\overline{P_{\rm g}}$ (panel c) and ($\rho$\,--\,$\overline{\rho})$\,/\,$\overline{\rho}$ (panel d) stratification {\it vs.} $\log\tau_{5000}$. Sequence from the centre of a granule (black colour, negative velocities) to the central part of its surrounding intergranular lane (red colour, positive velocities). Colours in between show the gradual granule-intergranular lane transition. Panel e) to h): Corresponding error bars presented for the centre of the granule and the central part of its surrounding intergranular lane.} 
\label{clust1}
\end{figure}

Figure \ref{clust1} shows the resulting granule to intergranular lane $T$\,--\,$\overline{T}$, $V_{\rm z}$\,--\,$\overline{V_{\rm z}}$, $(P_{\rm g}$\,--\,$\overline{P_{\rm g}})$\,/\,$\overline{P_{\rm g}}$ and ($\rho$\,--\,$\overline{\rho})$\,/\,$\overline{\rho}$ stratification {\it vs.} $\log\tau$. The presented error bars are calculated by the inversion code for each physical quantity and individual model using the response functions (for a recent explanation and application of this concept see Socas-Navarro \cite{socasnavarro04}). The error was propagated by assuming a Gaussian distribution until the values presented in panel e) to h). The error is minimal where the response functions (i.e. the sensitivities) are maximal for the spectral lines under study. The presentation of the physical quantities in panel a) to d) has been restricted to those layers with a reasonable degree of confidence.

In optical depth, the absolute temperature fluctuations $\Delta\,T$ (granule\,--\,intergranular lane) of the average granular cell (panel a) are rapidly decreasing from a maximum $\Delta\,T$ $\sim$\,400\,K (at $\log\tau$\,=\,0.2) towards a minimum near $\log\tau$\,=\,$-$\,1, and then increase again in the higher layers. At about $\log\tau$\,=\,$-$\,1.5 the granule appears cooler than the intergranular lane, thus confirming an inversion of the temperature contrast in higher layers, which is in agreement with our results of global correlation and spatial coherence and phase shift analyses presented in Paper I and with the results presented in Section \ref{cohphase}. It confirms other results reported in the literature (e.g. Deubner \cite{deubner88}; Collados et al. \cite{collados}; Rodr\'{\i}guez Hidalgo et al. \cite{rodriguez96}, \cite{rodriguez99}) and it has been predicted by theoretical models (Steffen et al. \cite{steffen89}; Stein \& Nordlund \cite{stein89}, \cite{steinnordlund98}; Gadun et al. \cite{gadun97}, \cite{gadun99}).

The absolute velocity fluctuations ($\Delta\,{V_{\rm z}}$), presented in Fig.  \ref{clust1} (panel b), reach 1.4\,km\,s$^{-1}$ at $\log\tau$\,=\,$-$\,0.4 decreasing slowly with height and are $\sim$\,0 at $\log\tau$\,=\,$-$\,3. Near the surface layers, velocities of the coolest intergranular lane are higher than the hottest granule ones.

\begin{figure}[]
\centering
\hspace{-0.15cm}
\includegraphics[width=8.5cm]{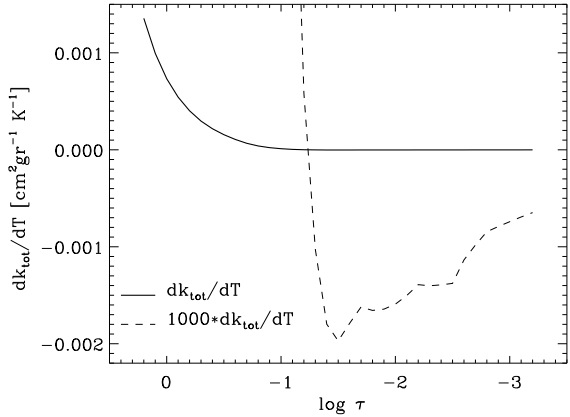}

\caption []{Derivative of the total continuum absorption coefficient ($k_{tot}$) {\it vs.} $T$, imposing hydrostatic equilibrium (HE).} 
\label{derivopacity}
\end{figure}

The relative pressure difference between the granule and the intergranular lane ($\Delta\,{P_{\rm g}}/{P_{\rm g}}$) decreases rapidly with decreasing optical depth and becomes practically 0 at $\log\tau$\,$\sim$\,$-$\,1 (see Fig. \ref{clust1}, panel c). The hydrostatic equilibrium equation on the optical depth scale reads:
\begin{equation}
\frac{d{P_{\rm g}}}{d \tau}=\frac{g}{k_{\rm tot}}
\label{HE_TAU}
\end{equation}
where $g$ stands for gravity and $k_{\rm tot}$ stands for total continuum absorption coefficient per mass unit. We see that the gas pressure depends only on the absorption coefficient per mass unit. In the Sun $k_{\rm tot}$ is dominated by $H^{-}$ opacity which, under the HE constraint, grows with $T$ in lower layers and slightly diminishes at higher ones (see Fig. \ref{derivopacity}). 
The derivative of $k_{tot}$ {\it vs.} $T$ has been calculated for the average model. To evaluate the derivative, we introduced a small $T$-perturbation and recalculated the $P_{\rm g}$-stratification and the absorption coefficients, imposing hydrostatic equilibrium (HE). Since granules are hotter at deeper layers the corresponding $k_{\rm tot}$ is larger and following Eq. \ref{HE_TAU} at larger optical depths the gas pressure in granules becomes smaller compared with intergranular lanes. At the higher layers (beyond $\log\tau$\,$\sim$\,$-$\,1) small positive values of $\Delta\,{P_{\rm g}}/{P_{\rm g}}$ are observed as a consequence of the changing sign of both the $T$-fluctuation and the derivative of $k_{\rm tot}$ (see Fig. \ref{clust1} and Fig. \ref{derivopacity}).

\begin{figure}[]
\centering
\hspace{-0.15cm}
\includegraphics[width=8.5cm]{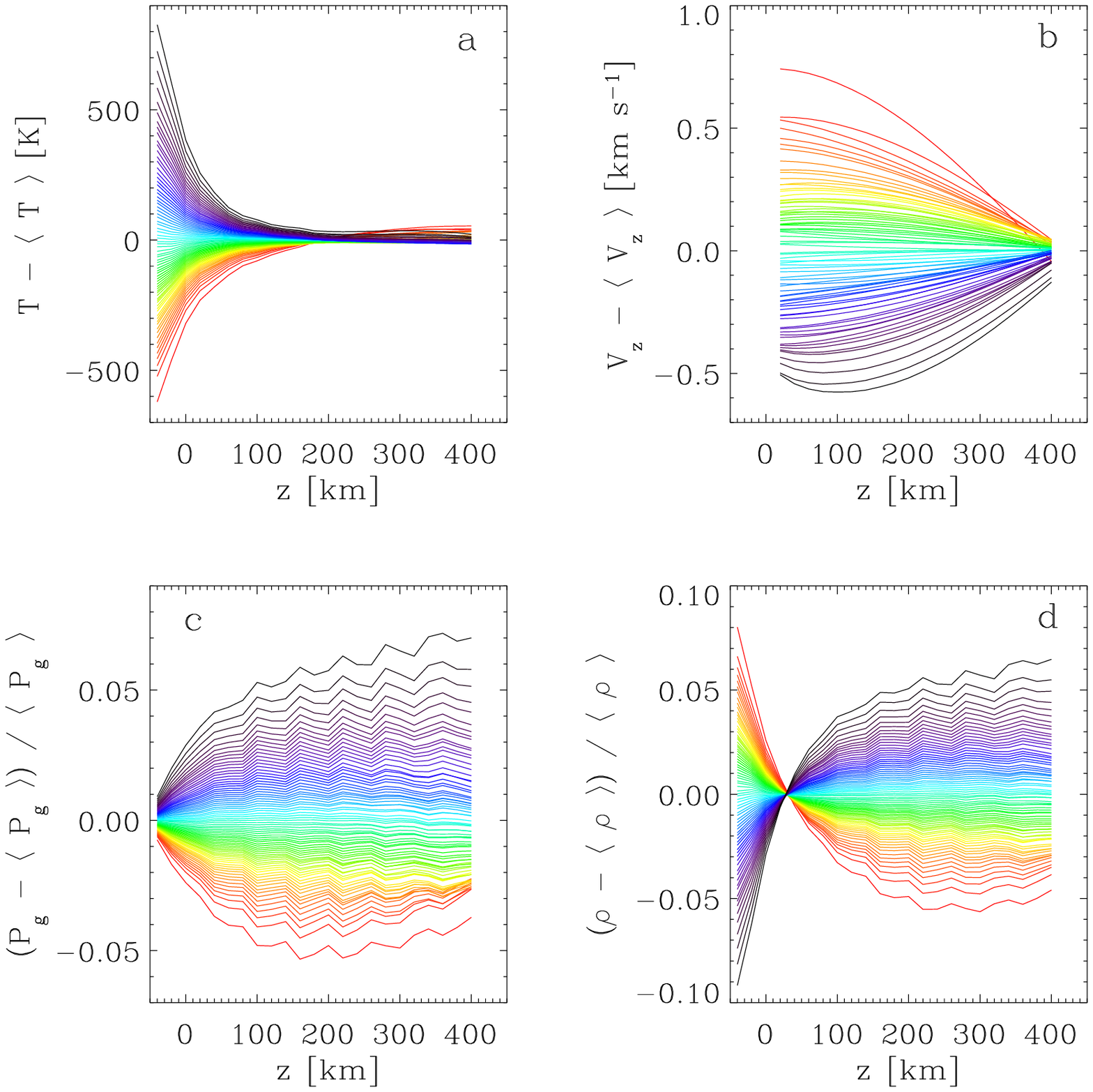}
\includegraphics[width=8.5cm]{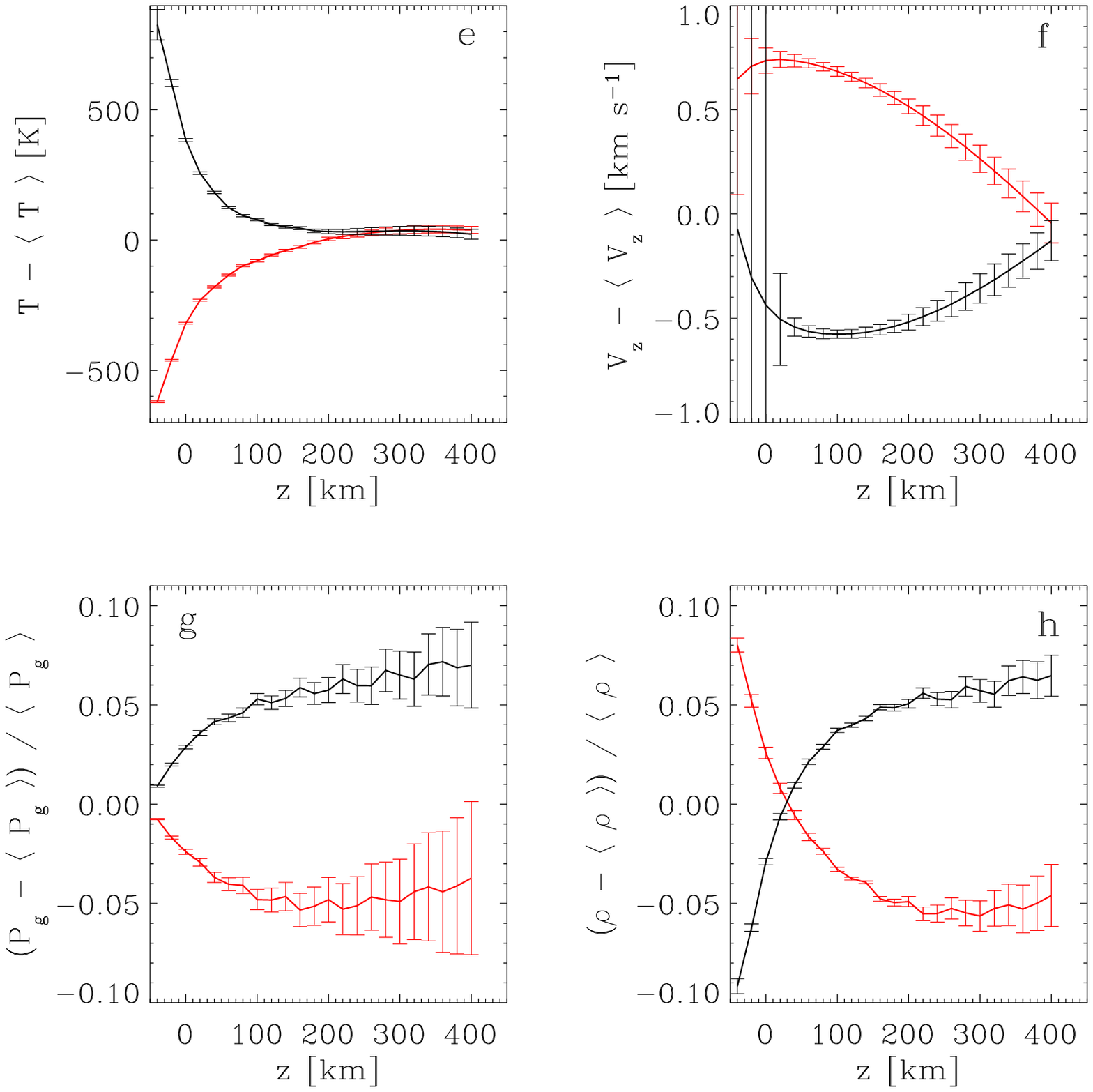}
\caption []{Model of an average granular cell: $T$\,--\,$\overline{T}$ (panel a), $V_{\rm z}$\,--\,$\overline{V_{\rm z}}$ (panel b), $(P_{\rm g}$\,--\,$\overline{P_{\rm g}})$\,/\,$\overline{P_{\rm g}}$ (panel c), $(\rho$\,--\,$\overline{\rho})$\,/\,$\overline{\rho}$ (panel d) stratification {\it vs.} $z$. Sequence from the centre of a granule (black colour, negative velocities) to the central part of its surrounding intergranular lane (red colour, positive velocities). Colours in between show the gradual granule-intergranular lane transition. Panel e) to h): Corresponding error bars presented for the centre of the granule and the central part of its surrounding intergranular lane.}
\label{clust2}
\end{figure}

The relative density fluctuations diminish faster than those of temperature but slower than the pressure ones. Considering the centre of a granule and calculating the scale height of $\Delta\,T /T$ (hereafter $H_{T}$), $\Delta\,{P_{\rm g}}/P_{\rm g}$ ($H_{P}$), and  $\Delta\,\rho /\rho$ ($H_{\rho}$) at the deepest layers, i.e. the interval in $\log\tau$ where the relative differences of these parameters is lower by a factor of $e$, we obtain values of $H_{T}$\,$=$\,0.76, $H_{P}$\,$=$\,0.48, and $H_{\rho}$\,$=$\,0.54. The intermediate behaviour of density can be explained in terms of the equation of state taking into account that the mean molecular weight ($\mu$) practically does not change at all. Hence at the layers where $\Delta\,{P_{\rm g}}/P_{\rm g}$ is vanishing $ \Delta\,\rho /\rho \sim - \Delta\,T /T$.

\subsection{Results as a function of geometrical height}
\label{modavz}

\begin{figure}[]
\centering
\hspace{-0.15cm}
\includegraphics[width=8.5cm]{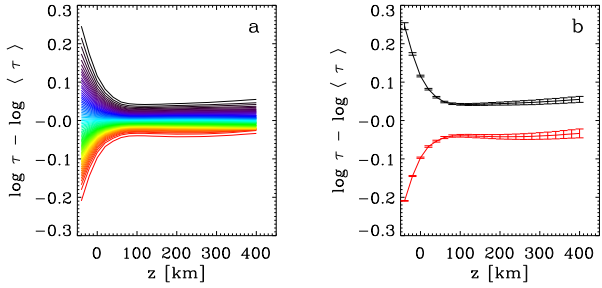}
\caption []{Model of an average granular cell. Panel a): $\log\tau$\,--\,$\log\overline{\tau}$ stratification {\it vs.} $z$. Sequence from the centre of a granule (black colour, negative velocities) to the central part of its surrounding intergranular lane (red colour, positive velocities). Colours in between show the gradual transition from a granule to its surrounding intergranular lane. Panel b): Corresponding error bars presented for the centre of the granule and the central part of its surrounding intergranular lane.}
\label{clust3}
\end{figure}

In Fig. \ref{clust2} the resulting granule to intergranular lane $T$\,--\,$\overline{T}$, $V_{\rm z}$\,--\,$\overline{V_{\rm z}}$, $(P_{\rm g}$\,--\,$\overline{P_{\rm g}})$\,/\,$\overline{P_{\rm g}}$ and ($\rho$\,--\,$\overline{\rho})$\,/\,$\overline{\rho}$ stratifications {\it vs.} $z$ are represented. The error bars have been calculated by a Monte-Carlo simulation, e.g. by adding random noise to each physical quantity of the individual models in $\log\tau$ with a $\sigma$ equal to the estimated error. For each model, perturbed in this way, the corresponding stratification in $z$ is evaluated. By repeating this process in a large number of simulations, the resulting error is given by the deviation with respect to the mean of the obtained results.

In geometrical height, the maximum $\Delta T$ increases significantly up to 1400\,K at z\,=\,$-$\,40\,km (compared with the 400\,K at $\log\tau$\,=\,0.2 in optical depth). This is in agreement with the results found by Rodr\'{\i}guez Hidalgo et al. (\cite{rodriguez99}). Also according to numerical simulations by Stein \& Nordlund (\cite{steinnordlund98}) there should be a much wider spread in temperatures at a given geometric height (near the surface) than there is on a local optical depth scale. Near the surface the energy transport switches from convective below the surface to radiative above the surface. The temperature gradient on an optical depth scale around $\log\tau$\,=\,0 corresponds to an extremely steep gradient on a geometric height scale because of the extreme temperature sensitivity of the dominant $H^{-}$ opacity, so that a small increase in temperature produces a large increase in opacity and hence a large increase in optical depth over a very small geometric height range. However, we obtain in geometrical height still a much smaller $T$ contrast compared with observed continuum intensity maps since our analyses are based on a time series of one dimensional slit spectra and thus we are not able to correct the data for the influence of degradation by stray light contamination. We find the $T$ fluctuations decrease rapidly with height and reaching a minimum at $\sim$\,170\,km, close to the value obtained by Rodr\'{\i}guez Hidalgo et al. (\cite{rodriguez99}), who localised a vanishing of $\Delta T$ at 150\,km. In agreement with the results obtained from the coherence and phase shift analysis (Fig. \ref{cohpha}), a weak inversion of the $T$ contrast at higher layers is found. 

Concerning vertical velocities, the maximum $\Delta V_{\rm z}$ amounts 1.4\,km\,s$^{-1}$ at $z$\,=\,60\,km (close to the value found for optical depth at $\log\tau$\,=\,$-$0.4). $\Delta V_{\rm z}$ decreases slowly from there on with increasing height, amounting to $\sim$\,0.4\,km\,s$^{-1}$ at $z$\,=\,400\,km. This is also in agreement with results by Rodr\'{\i}guez Hidalgo et al. (\cite{rodriguez99}), who found a value of $\sim$\,1\,km\,s$^{-1}$ at $z$\,=\,150\,km and very slowly decreasing afterwards. Initially hotter and less dense up-flowing plasma continues to rise even beyond the density inversion at $z$\,$\sim$\,30\,km, which clearly indicates observational evidence for substantial overshoot into the photosphere. As observed already in optical depth, the velocities of the coolest intergranular lane are higher than the hottest granule ones near surface layers, in concordance with numerical simulations by Stein \& Nordlund (\cite{steinnordlund98}). Also, panel f) depicts clearly that the lines are insensitive to velocities at the deepest layers. The error of vertical velocity  increases dramatically for $z$\,$<$\,20 km, thus the representation and analysis of velocity has been restricted to layers beyond. Nevertheless, by looking at the reliable layers, the granular and intergranular velocities show extrema (in absolute values) at $z$\,$\sim$\,80\,km and $z$\,$\sim$\,20\,km, the one of granular velocities shifted towards higher layers. This is in qualitatively good agreement with the simulations by Stein \& Nordlund (\cite{steinnordlund98}, Fig. 5).

The $\Delta P_{\rm g}$ values obtained (Fig. \ref{clust2}) could be strongly influenced by the criterion we have used to find a common geometrical height grid, where arbitrarily a layer of constant gas pressure was imposed. The exponential decrease of $P_{\rm g}$ and the fact that the layer of equal $P_{\rm g}$ has been selected deep in the photosphere (at z\,=\,$-$80\,km) can force spurious pressure differences from local hydrostatic equilibrium at some heights in our model. This problem will be addressed in a forthcoming paper by creating a dynamic model imposing the equations of continuity and motion. However, we can assume that temperatures and velocities are only weakly affected by this problem, showing a much smaller height dependence than pressure fluctuations. Densities again could be strongly affected by the changes introduced to  correctly solve the equations of motion. The corresponding panels of Fig. \ref{clust2} have to be treated carefully.

The decrease of the optical depth range (hereafter $\Delta\,\log\tau$) with geometrical height implies that granules are much less transparent than intergranular lanes in layers below 50\,km and only slightly less transparent in layers above, see Fig. \ref{clust3} where the $\log\tau$\,--\,$\log\overline{\tau}$ stratification is presented. Greater opacity for granules is observed in all layers which can be interpreted as follows. At layers where the granular temperature is lower than the average, the derivative of the continuum absorption coefficient is also negative (see Fig. \ref{derivopacity}).

 We suggest that this change of opacity enhances somewhat the crossing of temperatures (inversion of temperature contrast) between granules and intergranular lanes around $\log\tau$\,=\,$-$1 (see Fig. \ref{clust1}).

\section {Detailed analysis of the spatial variations of physical quantities across a small and a large granular cell}

\begin{figure}
\centering
\includegraphics[width=8.5cm]{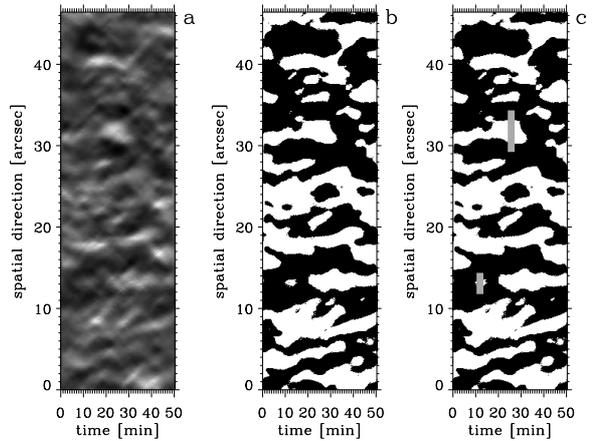}
\caption []{$T(t,x,\log\tau=0)$ image (panel a), corresponding $T$ binary image (panel b), and $T$ binary image showing the selected areas (marked by rectangles) for the study of the variations of physical parameters across a specific small and a specific large granular cell (panel c).}
\label{selectsize}
\end{figure}

\begin{figure*}[t]
\centering
\includegraphics[width=8.5cm]{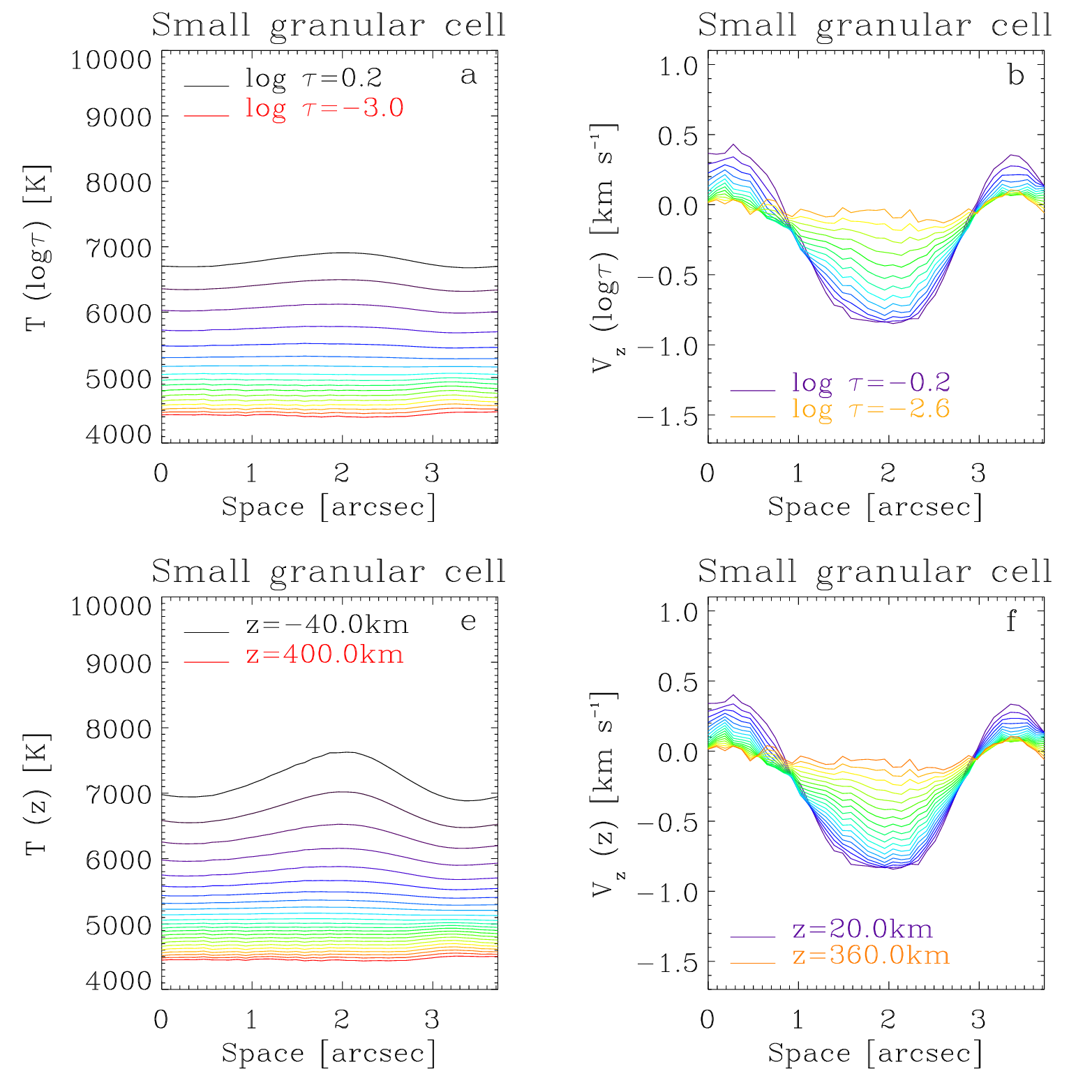}
\includegraphics[width=8.5cm]{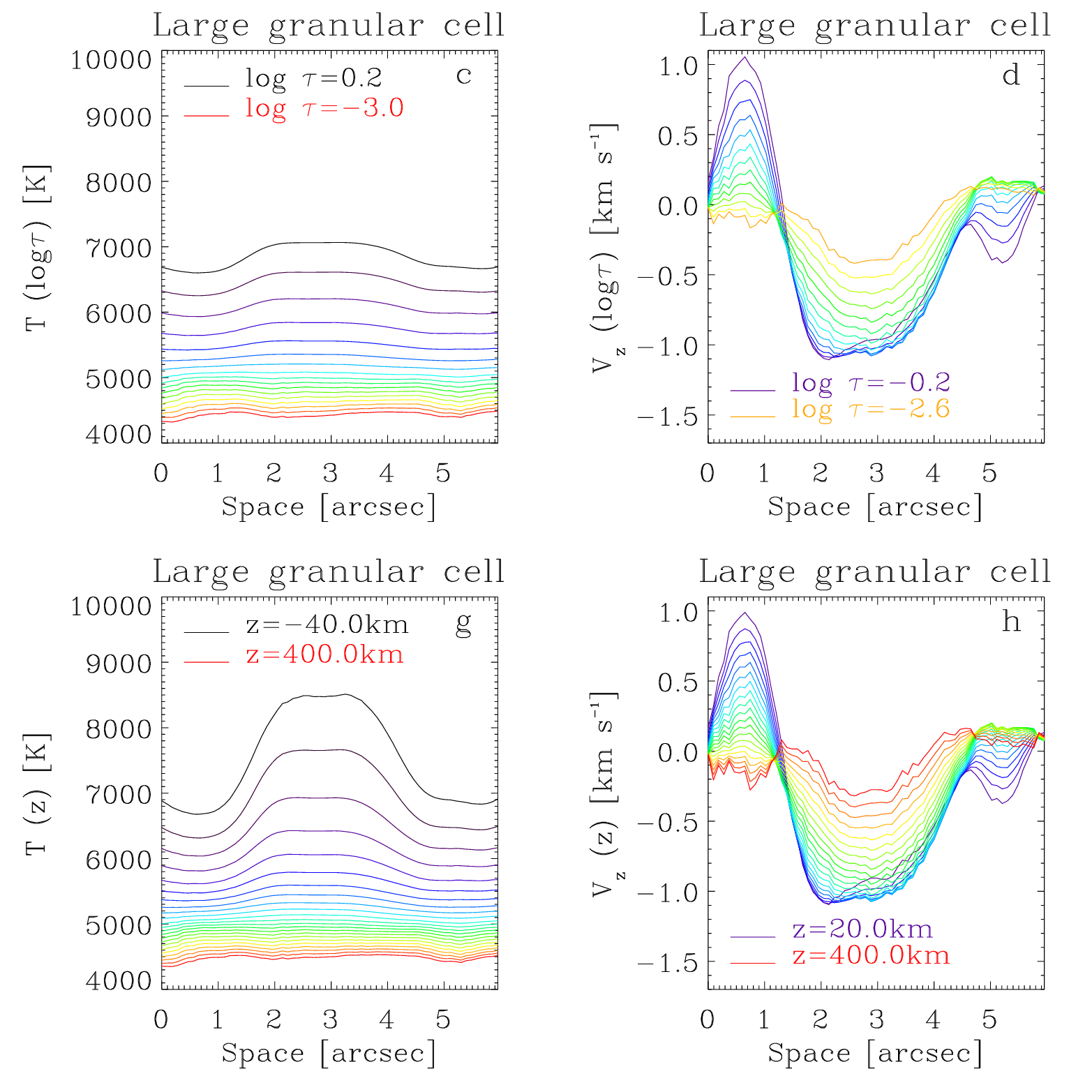}
\caption []{Panel a) to d): Spatial studies of the variation of $T$ and $V_{\rm z}$ across a particular small and large granular cell at different optical depths. Panel e) to h):  The same for geometrical heights. Different colours (sequence from black to red) are indicating different depths (heights). $T$ is presented from $\log\tau$\,=\,0.2 ($z$\,=\,$-$\,40\,km) up to $\log\tau$\,=\,$-$\,3 ($z$\,=\,$400$\,km) , $V_{\rm z}$ from $\log\tau$\,=\,$-$\,0.2 ($z$\,=\,20\,km) up to $\log\tau$\,=\,$-$\,2.6 ($z$\,=\,$360$\,km and $z$\,=\,$400$\,km).}
\label{space}
\end{figure*}

For a comparison with the more general results obtained by means of studies of the average granular cell and the spatial coherence and phase spectra analysis, the spatial variations of physical quantities, i.e. for temperature and vertical velocity, across a small (size $<$\,1\farcs5) and a large (size $\sim$\,3\arcsec) granular cell (see Fig. \ref{selectsize} for the selected areas) at different optical depths (geometrical heights) are presented in Fig. \ref{space} as temporal averages. 

Granular cells have been identified by means of a binary map of temperature at the $\log\tau$\,=\,0 level, which has been created by applying a criterion similar to the intensity criterion for the selection of granular and intergranular regions, described in Paper 1. Granules (intergranular lanes) are defined by all those pixels in space ($x$) and time ($t$) with temperature values at $\log\tau$\,=\,0, satisfying the following conditions:\\

\noindent{\it Granules:}

\begin{equation}
T(x,t,\log\tau=0)>\overline {T}(x,t,\log\tau=0)
\label{gran}
\end{equation}

\noindent{\it Intergranular lanes:}

\begin{equation}
T(x,t,\log\tau=0)<\overline {T}(x,t,\log\tau=0)
\label{ingran}
\end{equation}

\noindent where $\overline {T}$ is the mean temperature value over the whole image. \\

We are dealing with time slices, thus it is difficult to really identify granular cells in the 'images', i.e. a small bright point in the $x$\,-\,$t$ image could be also a part of a larger granule moving in and out of the slit. Likewise, it is difficult to cut through the centre of a large granule. The results presented here serve only for a qualitative comparison with the results obtained for the average cell. Analyses based on real 2D spectrometric data will provide more detailed insight in the future. By keeping the limitations in mind the main results can be summarised as follows:

The high temperature sensitivity of the $H^-$ opacity produces much smaller temperature fluctuations at a certain level in optical depth than in geometrical height (seen also in the case of the average granular cell in Figs. \ref{clust1} and \ref{clust2}). E.g., for the large cell in deep layers a $\Delta T$ of $\sim$\,1800\,K is observed, whereas in optical depth only a $\Delta T$ of $\sim$\,500\,K is detected. 

In general, we find the variation of physical quantities much more pronounced across the large granular cell, which is also in agreement with the results obtained in Paper I by means of a detailed analysis of the variation of line parameters in space and time for a specific small and a specific large granular cell. In the case of the small cell at deep layers a maximum $\Delta T$ of $\sim$\,200\,K in optical depth and of $\sim$\,800\,K in geometrical height is found. For the larger cell a very broad $T$ maximum at the centre of the granule is found, also in agreement with the results of Paper I. $\Delta T$ decreases rapidly from its maximum of 500\,K with decreasing optical depth (i.e. from 1800\,K with increasing geometric height). The decrease occurs even faster in the small cell.

For vertical velocities, as in the case of the average model (see Figs. \ref{clust1} and \ref{clust2}), we find similar values in optical depth and geometrical height. For the small cell, the maximum $\Delta V_{\rm z}$ amounts to $\sim$\,1.4\,km\,s$^{-1}$, whereas for the larger cell a maximum $\Delta V_{\rm z}$ of $\sim$\,2\,km\,s$^{-1}$ is found. 

In contrast to the average case (see Fig. \ref{clust2}) no larger velocities are found at the intergranular lane as compared with the granule. Note that in Fig. \ref{clust2} the largest downflows appear only for the coolest bin, which do not coincide with the samples presented here.

We also point out an asymmetric distribution of velocities in particular cases, e.g. at deep layers of the larger cell, the maximum up-flow is located near the granular border whereas at higher layers and in the case of the smaller cell the vertical speeds show increased blue shifts towards the granule centre. 

This is in agreement with the results obtained in Paper I by comparing the line core velocities of  the two \ion{Fe}{i} lines $\lambda$ 6494.98 \AA\, and $\lambda$ 6496.47 \AA. We conclude that in particular cases of very large granules an asymmetric distribution of vertical velocities across the cell can be observed. This is in agreement with the findings of other authors, e.g. de Boer et al. (\cite{deboer}), Nesis et al. (\cite{nesis93}), Hanslmeier et al. (\cite{hanslmeier94}), Hirzberger (\cite{hirzberger02}), indicating that maxima of up-flows in large granules do not always coincide with the maxima of intensity, and velocity peaks adjacent to the intergranular lanes can be observed.

\section {Conclusions}
The vertical structure of the solar photosphere has been investigated by means of different approaches. In Paper I, an analysis of global correlations, spatial coherence and phase spectra between different line parameters, obtained from a time series of 1D spectrograms, has been carried out. In the present work the spatial and temporal distribution of the thermodynamical quantities and the vertical flow velocity is derived as a function of optical depth ($\log\tau$) and geometrical height ($z$), applying the inversion technique SIR (Stokes Inversion based on Response functions) on the above data set. Spatial coherence and phase spectra between the fluctuations of temperatures and vertical velocities {\it vs.} optical depth and geometrical height have been computed for unfiltered and filtered data to determine the height variation of physical quantities of structures with different size. A model of an average granular cell has been derived, showing the granule--intergranular lane stratification of different parameters as a function of $\log\tau$ and $z$.  The cases of a specific small and a specific large granular cell have been investigated to compare with the results obtained in the more general cases. 

Regarding the granular temperature pattern, a fast decay of the temperature fluctuations between granules and intergranular lanes ($\Delta T$) with increasing height can be observed. A less efficient penetration of smaller cells to higher layers is reflected and already at $\sim$\,$\log\tau$\,=\,$-$\,1 or $z$\,$\sim$\,170\,km the $T$\,-\,$T$ coherence is lost at all granular scales and $\Delta T$ shows a minimum. At the layers beyond, for structures $>$\,1\farcs5 an inversion of the temperature contrast in both optical depth and geometrical height is found. We show how the effect of opacity somewhat enlarges, in the $\log\tau$ scale, the inversion of the temperature contrast, which has been ascribed so far only to a cooling produced by adiabatic expansion. Furthermore, in agreement with the numerical simulations by Stein \& Nordlund (\cite{steinnordlund98}), the high temperature sensitivity of the $H^-$ opacity produces at equal optical depths much smaller temperature fluctuations (we report a $\Delta T$ of 400 K at $\log\tau$\,=\,0.2) than at equal geometrical levels (where we find a $\Delta T$ of 1400 K at $z$\,=\,$-$\,40\,km). A maximum $\Delta T$ of 200\,K (800\,K) in optical depth (geometrical height) for a specific small granular cell and of about 500\,K (1800\,K) for the larger one, decaying rapidly with height, confirm the results obtained for the average cell.

A much slower decay with height of the vertical convective velocity fluctuations between granules and intergranular lanes ($\Delta V_{\rm z}$) is found. In deep photospheric layers, a maximum $\Delta V_{\rm z}$ of 1.4\,km\,s$^{-1}$ in both optical depth and geometrical height is in concordance with the 1.2\,km\,s$^{-1}$ and 2\,km\,s$^{-1}$ found for the specific small and specific large granular cell. Going up in height, as in the case of temperature, a less efficient penetration of convective velocities of smaller cells is reflected, though even at $\log\tau$\,$\sim$\,$-$\,2.0 and $z$\,=\,280\,km structures $>$\,1\,\farcs4 are found. At $z$\,$\sim$\,370\,km, only velocities of structures at the largest granular scales ($\sim$\,4\arcsec) are present. This similar size distribution between velocities and temperatures with height together with the temperature inversion clearly provides observational evidence for substantial overshoot into the photosphere, far beyond the density inversion, which we find near the surface layers. At deep photospheric layers, the behaviour of the vertical velocities reflected in the simulations of Stein \& Nordlund (\cite{steinnordlund98}) is for the first time qualitatively reproduced by observations. In our model the centre of the granule shows smaller velocities than the centre part of the intergranular lane. Both velocities (in absolute values) reach extrema near the surface layers, but the granular one is shifted towards higher layers. Concerning the specific studies of a small and large granular cell we point out an asymmetric distribution of vertical velocities for the large granular cell at the deep layers, where the maximum up-flow is located near the granular borders. However, at higher layers and for the specific small cell the vertical speed shows a more symmetric distribution with increased blue shift towards the granule centre. This confirms findings of other authors, e.g. Hirzberger (\cite{hirzberger02}), indicating that maxima of up-flows in large granules do not always coincide with the maxima of intensity, and velocity peaks adjacent to the intergranular lanes can be observed.

The obtained gas pressure differences between granules and intergranular lanes ($\Delta P_{\rm g}$) could be influenced by the criterion we have used to find a common geometrical height grid, defining arbitrarily a layer with horizontally equal $P_{\rm g}$. A deep photospheric layer has been selected for the horizontal pressure equilibrium. This together with the exponential decrease of  $P_{\rm g}$ can introduce spurious pressure differences from hydrostatic equilibrium at some heights in our model. This problem will be addressed in the next paper of this series, by imposing the equations of continuity and motion to our model, thus creating a 2D semi-empiric dynamic model of an average granular cell. 

\acknowledgements{We thank the referee, Dr. M. Steffen, for suggestions and comments that have significantly improved this paper. We are greatful for the support by the Austrian Fond zur F\"orderung der wissenschaftlichen Forschung (Proj. Nr. 9184-PHY). This work was partially supported by the Spanish DGES under project 95-0028 and the Austrian-Spanish "Acciones Integradas" (grant no. 97-0028). Partial support by the Spanish Ministerio de Ciencia y Tecnolog\'\i a and by FEDER through project AYA2001-1649 and by the Deutsche Forschungsgemeinschaft through grant KN 152/29-1 is gratefully acknowledged. The Vacuum Tower Telescope is operated by the Kiepenheuer-Institut f\"ur Sonnenphysik, Freiburg, at the Spanish Observatorio del Teide of the Instituto de Astrof\'\i sica de Canarias.}

%-----------------------------------

\end{document}